\documentclass[]{pasj02} 
\usepackage{ulem}

\jyear{2025}
\Received{2025/06/19}
\Accepted{2025/11/04}

\begin{document} 

\title{Structure and Magnetic Field of the Bright-rimmed Cloud IC~1396E/SFO~38 }

\author{
 Koji \textsc{Sugitani},\altaffilmark{1}\altemailmark
 \email{sugitani@nsc.nagoya-cu.ac.jp} 
 Jan~G.~A. \textsc{Wouterloot},\altaffilmark{2}\altemailmark
 Harriet \textsc{Parsons},\altaffilmark{2}\altemailmark
 Sarah \textsc{Graves}\altaffilmark{2}\altemailmark
 Takayoshi \textsc{Kusune}\altaffilmark{3}\altemailmark
and
 Archana \textsc{Soam}\altaffilmark{4}\altemailmark
}
\altaffiltext{1}{Graduate School of Science, Nagoya City University, Mizuho-ku, Nagoya, Aichi  467-8501, Japan}
\altaffiltext{2}{East Asian Observatory, 660 N. A'ohoku Place, Hilo, Hawaii 96720, USA}
\altaffiltext{3}{Department of Physics, Nagoya University, Chikusa-ku, Nagoya, Aichi 464-8602, Japan}
\altaffiltext{4}{Indian Institute of Astrophysics, II Block, Koramangala, Bengaluru 560034, India}




\maketitle

\begin{abstract}
We carried out polarimetric observations of the bright-rimmed cloud IC~1396E/SFO~38 with SCUBA-2/POL-2 to study the effect of ultraviolet (UV) light on its structure and magnetic field. 
This bright-rimmed cloud appears optically to be one single cloud illuminated by the UV light from the exciting star of IC 1396.  However our Stokes I image and $^{13}$CO~($J=3-2$) archival data suggest that this cloud is not a simple, single structure, but appears to be composed of two parts on first glance; a head part with wings and a tail, and a north-west extension part.
Since molecular clouds are generally filamentary and it seems likely that the initial structures of bright-rimmed clouds are expected to be also generally elongated, we examined the possibility that the structure was created from a single elongated cloud by the UV impact.
We compared the cloud structure with a simulation study that investigated the evolution of prolate clouds exposed to the UV radiation from various directions and found that this apparent two-part structure could be reproduced in a situation where a single filamentary cloud is obliquely illuminated by UV light.
The magnetic field directions of the cloud are different from the ambient field direction, demonstrating the field reconfiguration.
The distortion or pinch of the magnetic field is seen toward the cloud head, where an intermediate-mass star cluster is located, suggesting gravitational contraction.
We roughly estimated the magnetic strength and stability in three parts of the cloud and found that the cloud head is most likely to be supercritical.
\end{abstract}


\section{Introduction} \label{sec:intro}
 
The ultraviolet  (UV) radiation from OB stars ionizes and destroys their parent molecular clouds, creating  H\emissiontype{II} regions. On the other hand, it also compresses some dense parts of the molecular clouds and promote star formation there.
Bright-rimmed clouds (BRCs) are dense cloudlets, often seen at the periphery of extended H\emissiontype{II} regions. They have ionization fronts on the side of OB stars that excite H\emissiontype{II} regions and are considered to be potential sites for induced star formation by UV radiation from nearby OB stars (e.g., \cite{1998ASPC..148..150E,2011EAS....51...45E}, and references therein).  
Many BRCs are relatively isolated and this situation seems good for studying the UV impact on molecular clouds.

To understand dynamical evolution of such molecular clouds by so-called radiation-driven implosion (RDI) and star formation process there, many 2D/3D radiation-hydrodynamics calculations have been done (e.g., \cite{1982ApJ...260..183S,1994A&A...289..559L, 2001MNRAS.327..788W, 2003MNRAS.338..545K, 2006MNRAS.369..143M, 2009ApJ...692..382M, 2009MNRAS.393...21G, 2010MNRAS.403..714M, 2011ApJ...736..142B}).  
 Although these simulations did not include all possible physical effects that could impact on the cloud evolution (e.g., self-gravity, stellar feedback, initial perturbations, etc.), they succeeded in reproducing some of the observed properties of BRCs.
 
One such effect not always considered is the asymmetric, geometrical configuration of the initial cloud.
In most of the simulations,  (nearly) symmetrical configurations with respect to the incident UV radiation were adapted as the initial configuration, e.g., spherical distribution, pillar-like distribution parallel to the UV radiation. 
 \citet{2014MNRAS.444.1221K} and \citet{2015MNRAS.450.1017K}  conducted radiation-hydrodydamics calculations in order to investigate the prolate cloud evolution at H\emissiontype{II} boundaries. 
 Their trial makes sense, because many molecular clouds have been shown to be  elongated structures, i.e., filamentary.
\citet{2015MNRAS.450.1017K} used a set of four parameters of the simulated clouds (the number density, the prolate ratio of major to minor axis, the inclination angle between the cloud major axis and the UV incident direction, and the incident UV flux) and examined the dependency of these four parameters under RDI.
They found that many types of asymmetrical BRCs, including those with filamentary structures and irregular horse-head structures, could be developed from the initial prolate cloud and showed that their simulated structures resemble the appearances of BRCs in the optical images.
However, the density and/or velocity structures of BRCs have not been well investigated by observation in term of the asymmetry structures of BRCs, and  millimeter/sub-milimeter studies are needed. 

Another effect not always considered is the magnetic field in most astronomical simulations of BRCs.  
The magnetic field is believed to play an important role in most astronomical processes, but the magnetic effects are not always treated in these simulations for BRCs. 
 \citet{1989ApJ...346..735B} and \citet{1990ApJ...354..529B}  analytically studied the initial implosion and subsequent equilibrium cometary stages of BRCs by approximately considering the  magnetic field.
They suggested that the magnetic field is a significant factor of the cloud pressure and gravitational stability in the equilibrium stage.
\citet{2013ApJ...766...50M} carried out numerical modeling of one BRC by assuming its equilibrium cometary stage and found that its observed density structure cannot be explained without the magnetic field. 
\citet{2009MNRAS.398..157H} conducted the first 3D radiation-magnetohydrodynamic (MHD) simulations of the dense, magnetized molecular globules photoionized by strong UV radiation. 
Their results showed that strong magnetic fields (with magnetic pressure over 100 times the gas pressure) cause significant deviations from the non-magnetic evolution.
\citet{2011MNRAS.412.2079M} also conducted 3D radiation-MHD simulations to investigate the effects of initially uniform magnetic fields on the formation and evolution of elongated clouds, such as  the M~16 pillars, at the boundaries of H\emissiontype{II} regions.  
They suggested that the weak initial field is altered toward those elongations by the UV light during the cloud evolution, while the strong field remains its initial configuration.

To verify these predictions of modeling and simulations on the role of magnetic field, it is necessary to reveal the magnetic field structures and strength of BRCs by observation.
Magnetic fields can be revealed in the visible/near-infrared (IR) wavelengths by starlight polarization due to interstellar grain alignment based on radiative processes (e.g., \cite{2015ARA&A..53..501A}).  
In the far-IR/submillimeter wavelengths, the polarized thermal radiation from the aligned grains enable to probe magnetic fields, the direction of which is perpendicular to the polarization angle.

At visible wavelengths, polarimetric observations of cometary globules in the Gum-Vela region (CG~22, CG~30-31 complex, CG~12) have been done \citep{1996MNRAS.279.1191S, 1999MNRAS.308...40B, 2004MNRAS.348...83B}.
A few more polarimetric observations have been reported; e.g., SFO~20 \citep{2011ApJ...743...54T}, LBN~437 \citep{2013MNRAS.432.1502S}, and IC~63 and IC~59 \citep{2017MNRAS.465..559S}. 
These optical polarimetric observations have shown that the magnetic fields at low column densities seem to be along the tails or edges of BRCs, but could not probe the denser parts of BRCs.
In the near-IR wavelengths, a few polarimetric studies have been reported; e.g., M~16 \citep{2007PASJ...59..507S}, SFO~89 in Sh~2-29 \citep{2014ApJ...783....1S}, SFO~74 \citep{2015ApJ...798...60K}.
These near-IR observations could probe rather denser parts of BRCs, but not well enough to probe the densest parts where star formation would be expected.  
In the submillimeter range, \citet{2018ApJ...860L...6P} first presented high-resolution polarimetric observations of the M~16 pillars.
They confirmed the previous report that the magnetic fields of the pillars are aligned along their elongation and are different from the global field structure in M~16 \citep{2007PASJ...59..507S} and suggested that the pillars have been formed through the compression of weakly magnetized gas based on the field strength estimate. 

At present, the number of magnetic field observations is still small and not sufficient to clearly reveal the field structures of BRCs, especially for their high-density parts. 
Thus, more polarimetric observations need to be made of the high-density parts of BRCs.

\subsection{IC~1396E/SFO~38} \label{sec:intro-sub}

\begin{figure*}[ht!]
\begin{center}
\includegraphics[width=17.1cm]{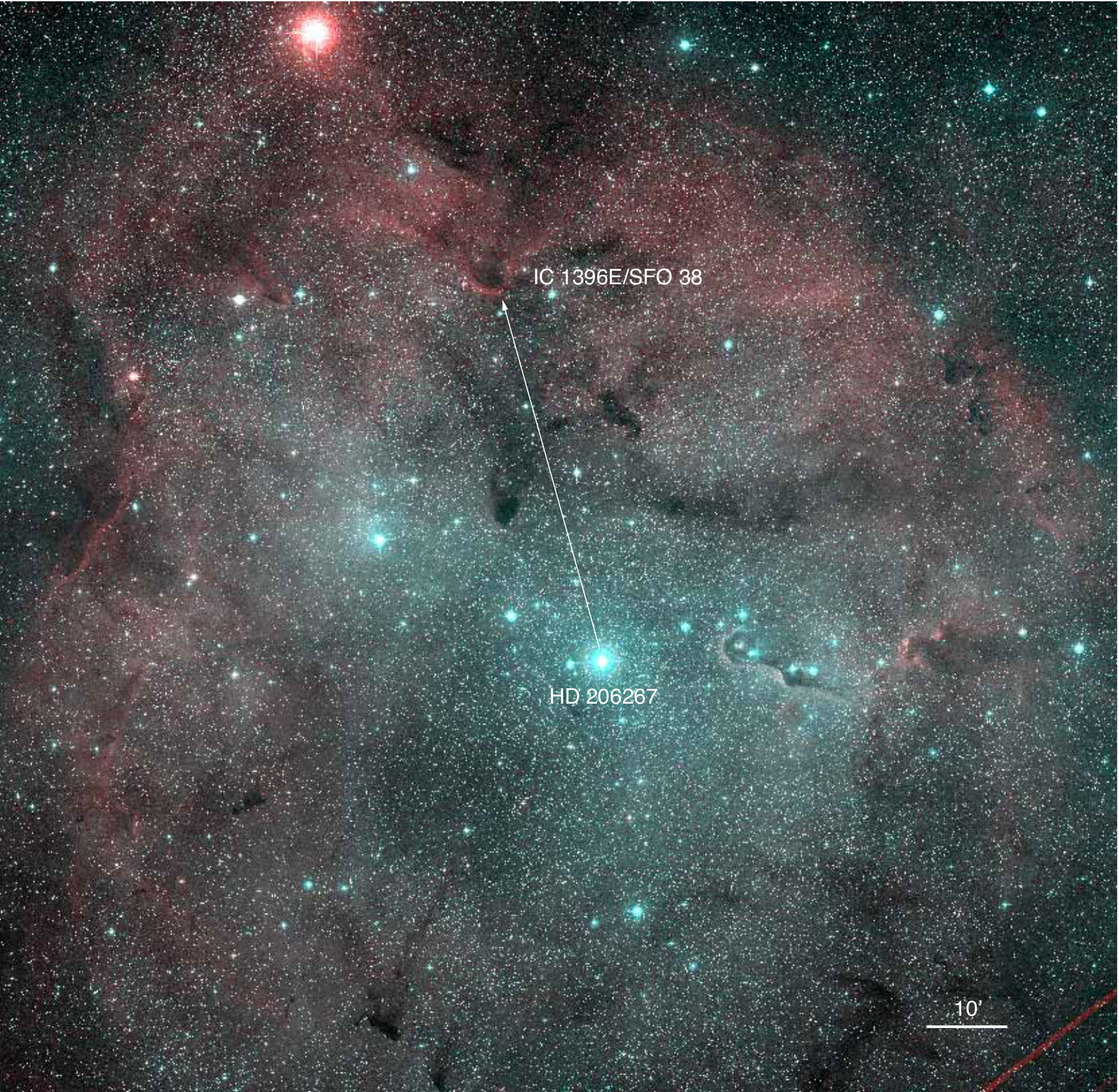}
\end{center}
\caption{False color image of IC~1396 (red: POSS2 red,  green and blue: POSS2 blue). 
North is at the top and east to the left. 
The UV incident direction from the main exciting star (HD~206267) of IC~1396 to IC~1396E/SFO~38 is indicated by an arrow.
}
\label{fig1}
\end{figure*}

IC~396E/SFO~38 is a BRC located toward the periphery of the H\emissiontype{II} region IC~1396 \citep{1956BAN....13...77P,1991ApJS...77...59S}. 
This H\emissiontype{II} region is a ring-shaped giant H\emissiontype{II} region in Cep OB2 with a size of more than $2 \arcdeg$ (see figure  \ref{fig1} ). 
The main exciting star is HD~206267, which is a spectroscopic binary of O6 and O9 \citep{1975AJ.....80..454C,1984ApJ...286..718W}. 
There are many bright-rimmed globules \citep{1956BAN....13...77P} in and around IC~1396 and molecular cloudlets,  including the globules, were mapped in the rotational transitions of CO and its isotopes \citep{1995ApJ...447..721P, 1996A&A...309..581W}. 
 \citet{2002AJ....124.1585C} derived the photometric distance of IC~1396 to be $\sim870~\pm~$80 pc. 
Recently, \citet{2019A&A...622A.118S} derived $947_{-73}^{+90}$ pc from Gaia DR2 \citep{2018A&A...616A...1G} and, more recently, \citet{2021AJ....162..279S}  gave $931_{-116}^{+151}$
from Gaia EDR3 \citep{2021A&A...649A...1G}.
These estimated distances are around 900 pc and seem to be consistent.
Here, for the sake of simplicity, we adopt 900 pc for the distance to IC~1396.  

In IC~1396E/SFO~38, \citet{1989ApJ...342L..87S} discovered a prominent molecular outflow (IC1396N) associated with the IRAS point source 21391+5802 and subsequently more detailed outflow studies with single dishes and interferometers have been done in more detail (e.g., \cite{1990AJ....100..758W, 1993AJ....106..250W, 2001A&A...376..271C, 2002ApJ...573..246B, 2009A&A...507.1475F}). 
The dense part of the cloud was mapped in various molecular lines \citep{1993ApJ...404..247S, 2001A&A...376..271C,  2002aprm.conf..213S}. 
The bolometric luminosity of IRAS~21391+5802 was  estimated in the range from 235 $L_\odot$ to 440 $L_\odot$ (e.g., \cite{1993AJ....106..250W, 1996A&A...315L.293S, 2000AJ....119..323S}) on the assumption of  a distance of 750 pc \citep{1979A&A....75..345M}.
This value of a few hundred solar luminosities and the detection of water maser emission \citep{1992A&A...255..293F, 1995A&AS..112..299T, 1999ApJ...526..236S, 2000ApJ...538..268P}, as well as the prominent outflow, indicate that IC~1396E/SFO~38 is an ongoing star-forming site for an intermediate-mass star.
IRAS~21391+5802 was resolved into three sources (BIMA 1-3; \cite{2002ApJ...573..246B}), where BIMA 2 and 3 respectively correspond to Sources A and B of  \citet{2001A&A...376..271C}, and BIMA 2 was shown to be an intermediate-mass protostar. 
The structures and kinematics around the IRAS source were examined in detail \citep{2004A&A...426..941B, 2007A&A...468L..33N, 2009A&A...507.1475F}.
A number of 2.12~$\mu$m H$_2$ line emission knots and Herbig-Haro flows were detected and their exciting sources were found away from the IRAS source position \citep{2001A&A...376..553N, 2003ApJ...593L..47R, 2009A&A...504...97B, 2012A&A...542L..26B}, indicating that on-going star formation takes place not only around the IRAS  position but also in other places within the cloud. 

A spatial concentration of H$\alpha$ emission stars, candidates of pre-main-sequence stars (PMSs), just outside the rim of IC~1396E/SFO~38 on the HD~206267 side was shown first by \citet{2002AJ....123.2597O} and subsequently by \citet{2012AJ....143...61N}. 
They pointed out the possibility of triggered star formation of low-mass stars due to the UV radiation from HD~206267.
\citet{2007ApJ...654..316G} made X-ray observations and found the elongated spatial distribution of X-ray sources, for which an age gradient oriented toward HD
206267 was derived, with the aid of near-IR and mid-IR data, supporting the idea of triggered star formation. 
On the other hand, \citet{2009A&A...504...97B} pointed out that from their near-IR imaging no clear evidence for triggered star formation was found because of the non-detection of  near-IR excess for the sources outside the clouds.  
However, \citet{2010ApJ...717.1067C}  showed an evolutionary sequence of YSOs from the rim to the dense core of the cloud with the Spitzer data and their optical photometry and spectroscopy of YSOs and PMSs, again supporting the idea of triggered star formation.  
Although the triggered star formation here is a matter of debate, all observations to date indicate that the most recent star formation (i.e., Class 0/I sources and/or outflow powering sources) is taking place within the dense parts of the cloud  (e.g., BIMA 1-3, source C, source I,  \#252, and  \#331 in \cite{2012A&A...542L..26B}). 

\citet{2018MNRAS.476.4782S} measured the magnetic fields toward four BRCs in IC~1396 by optical polarimetry and showed that the field direction around IC~1396E/SFO~38 is at PA $\sim50\arcdeg$, where PA is an angle measured counterclockwise from the north.  
However, their optical polarimetry only probed the magnetic fields just outside the clouds, not inside. 
It would be very interesting to examine how the magnetic fields are inside and how those have been altered from the original configurations by the UV radiation.
So, we have measured the magnetic field of one of these BRCs, IC~1396E /SFO~38, with JCMT/POL-2 and studied the effect of the UV on the magnetic field as well as the density and velocity structures.

\section{Observations and Data Reduction} \label{sec:obs}

\subsection{POL-2/SCUBA-2} \label{sec:obs-pol2}

We have made 850 $\mu$m polarimetric observations of IC~1396E/SFO~38 with POL-2/SCUBA-2 mounted on JCMT, on 2019 June 13 and 14.  
13 sets of $\sim$31 minute integration were made and the atmospheric optical depths at 225~GHz were between 0.04 and 0.07 during the observations.
We fully sampled two circular regions 2$\arcmin$ apart in the north-south direction with a diameter of 12$\arcmin$ and a resolution of $14.\arcsec1$ using the SCUBA-2 DAISY mapping mode \citep{2013MNRAS.430.2513H}, which is optimized for POL-2 observations.  
The telescope scan speed of the POL-2 DAISY is 8$\arcsec$~s$^{-1}$, with a half-waveplate rotation speed of 2~Hz \citep{2016SPIE.9914E..03F}.  
In the POL-2 DAISY scan pattern, an area with a diameter of 3$\arcmin$ in the center of the 12$\arcmin$ field is almost uniformly covered, with the noise increasing toward the map edge.
We observed two adjacent overlapping circular regions to fully cover the dense region of the IC~1396E/SFO~38 cloud \citep{1993ApJ...404..247S, 2001A&A...376..271C,  2002aprm.conf..213S}. 
The 450~$\mu$m polarimetric data was obtained simultaneously, but not treated in this paper.
The output maps of Stokes $Q$, $U$, and $I$ are gridded to 4$\arcsec$ pixels. 
The Stokes $I$ map was first calibrated in Jy~arcsec$^{-2}$, by using the flux conversion factor FCF (850)$_{\rm arcsec}$ of 2.34~Jy~pW$^{-1}$~arcsec$^{-2}$ \citep{2013MNRAS.430.2534D} and a correction factor of 1.35 for losses from POL-2 (Tutorial 1 on the JCMT web site), then to Jy~sr$^{-1}$.
The observed data were reduced by using the $pol2map$ routine of SMURF \citep{2013MNRAS.430.2545C} with a binsize parameter of 12$\arcsec$ to derive Stokes $I$, polarization degree ($P$), polarization angle from North (PA  or $\theta$), following POL-2 Data Reduction in Tutorial 1 on the JCMT web site. 
The polarization degree is calculated as 
\begin{equation}
  P = \sqrt{Q^2+U^2-0.5(\Delta Q^2 + \Delta U^2)} / I \times 100 \%, 
\end{equation}  
assuming $\Delta Q \sim \Delta U$, and the polarization angle as 
\begin{equation}
  \theta = \frac{1}{2}\tan^{-1}\frac{U}{Q}. 
\end{equation}  
The detailed summary of the data reduction with the $pol2map$ routine can be seen, e.g., in section 3 of \citet{2018ApJ...859....4K}.

\subsection{Archival Data} \label{sec:arc}

The JCMT Science Archive at the CADC (Canadian Astronomy Data Centre) was used to obtain the processed data of $^{12}$CO, $^{13}$CO, C$^{18}$O ($J=3-2$) of IC~1396E/SFO~38. 
These data were used to reveal  the large-scale cloud structure, molecular outflow lobes, and velocity structure, respectively, and to estimate the cloud velocity dispersion.

The Spitzer 8~$\mu$m \citep{2004ApJS..154....1W} and the WISE 22~$\mu$m \citep{2010AJ....140.1868W} archival data were used to indirectly show the existence of ionization fronts associated with the cloud (see section \ref{subsec:cs}).

Far-IR (353 GHz) polarization data from the Planck satellite \citep{2015A&A...576A.104P} was used to examine the local magnetic field structure around IC~1396E/SFO~38.

\subsection{Near-Infrared Data} \label{sec:nir}

$K$s- and $H$-band photometric data were obtained with the SIRIUS camera \citep{1999sf99.proc..397N, 2003SPIE.4841..459N}  mounted on the 2.2-m telescope of University of Hawaii, during the SIRIUS commissioning run. 
These data were used for comparison with other data.

\section{Results} \label{sec:result}

\subsection{Cloud Structure}\label{subsec:cs}

\begin{figure*}[t]
\begin{center}
\includegraphics[width=16cm]{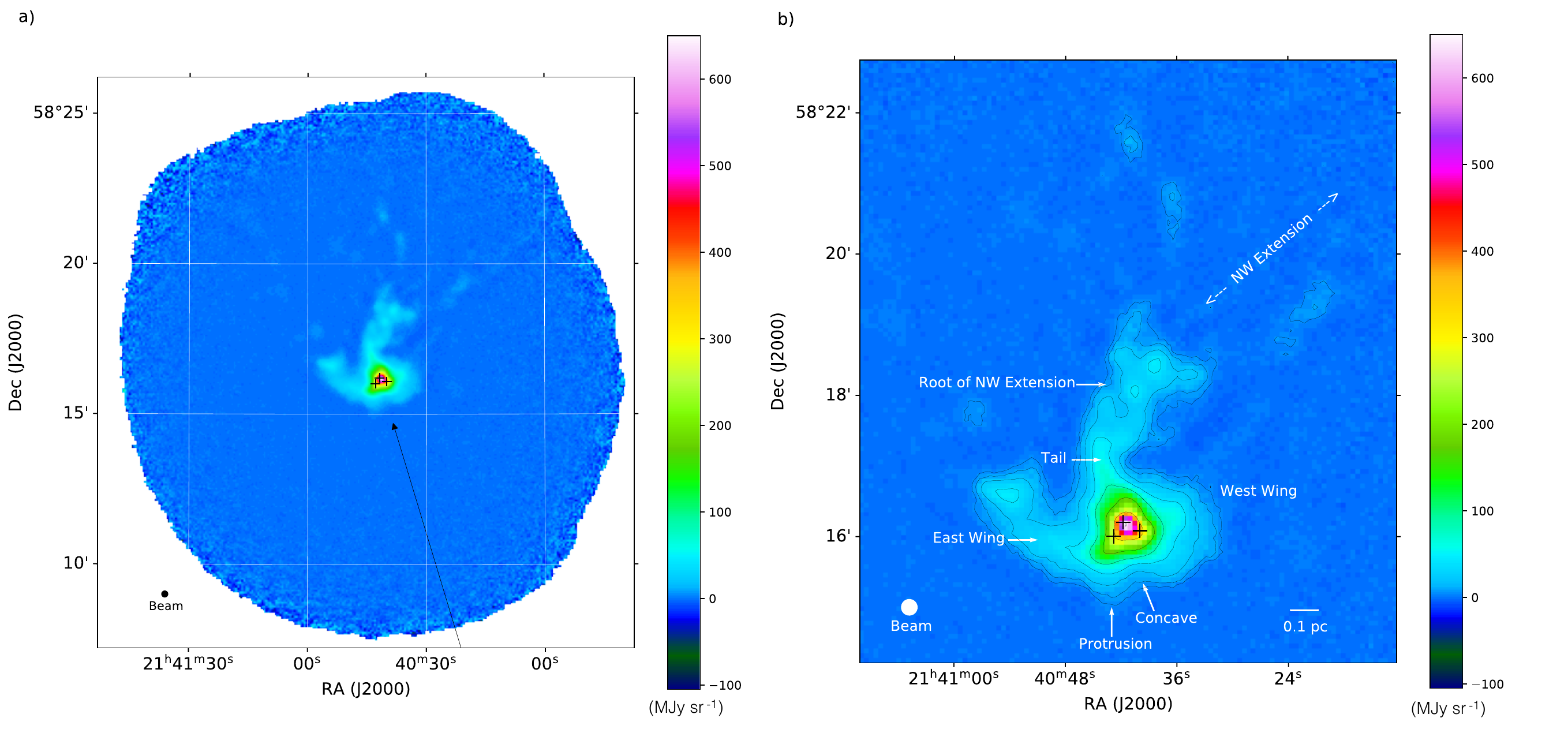}
\end{center}
\caption{Stokes I image of IC~1396E/SFO~38 obtained by the JCMT with SCUBA-2/POL-2.  
The beam size is $14 .\arcsec 1$, which corresponds to $\sim 0.062$ pc at the adopted distance of 900 pc. 
The intensity color scale in MJy~sr$^{-1}$ is shown on the right side of each image. 
The positions of the three continuum sources (BIMA 1-3; \cite{2002ApJ...573..246B}) are shown with plus marks.
(a) The UV incident direction from the main exciting star (HD~206267) of IC~1396 is indicated by an arrow.
(b) The enlargement of (a) with the contours of  5, 10, 20, 40, 80, 160, 320, and 640~MJy~ sr$^{-1}$.
}
\label{fig2}
\end{figure*}

Figure  \ref{fig2}  shows the Stokes $I$ map (850 $\mu$m intensity map)  obtained with POL-2/SCUBA-2.  
In this map, the areas with a rms noise level of $\lesssim$ 1.0~MJy~sr$^{-1}$ and $\lesssim$ 1.5 ~MJy~sr$^{-1}$   are roughly estimated to be $\sim3.5 \arcmin$ and $\sim5.0 \arcmin$ from the approximate map center, (RA,~Dec)$_{\rm J2000.0}$ = (\timeform{21h40m45s}, \timeform{58D17'3}).

\begin{figure*}[t]
\begin{center}
\includegraphics[width=15cm]{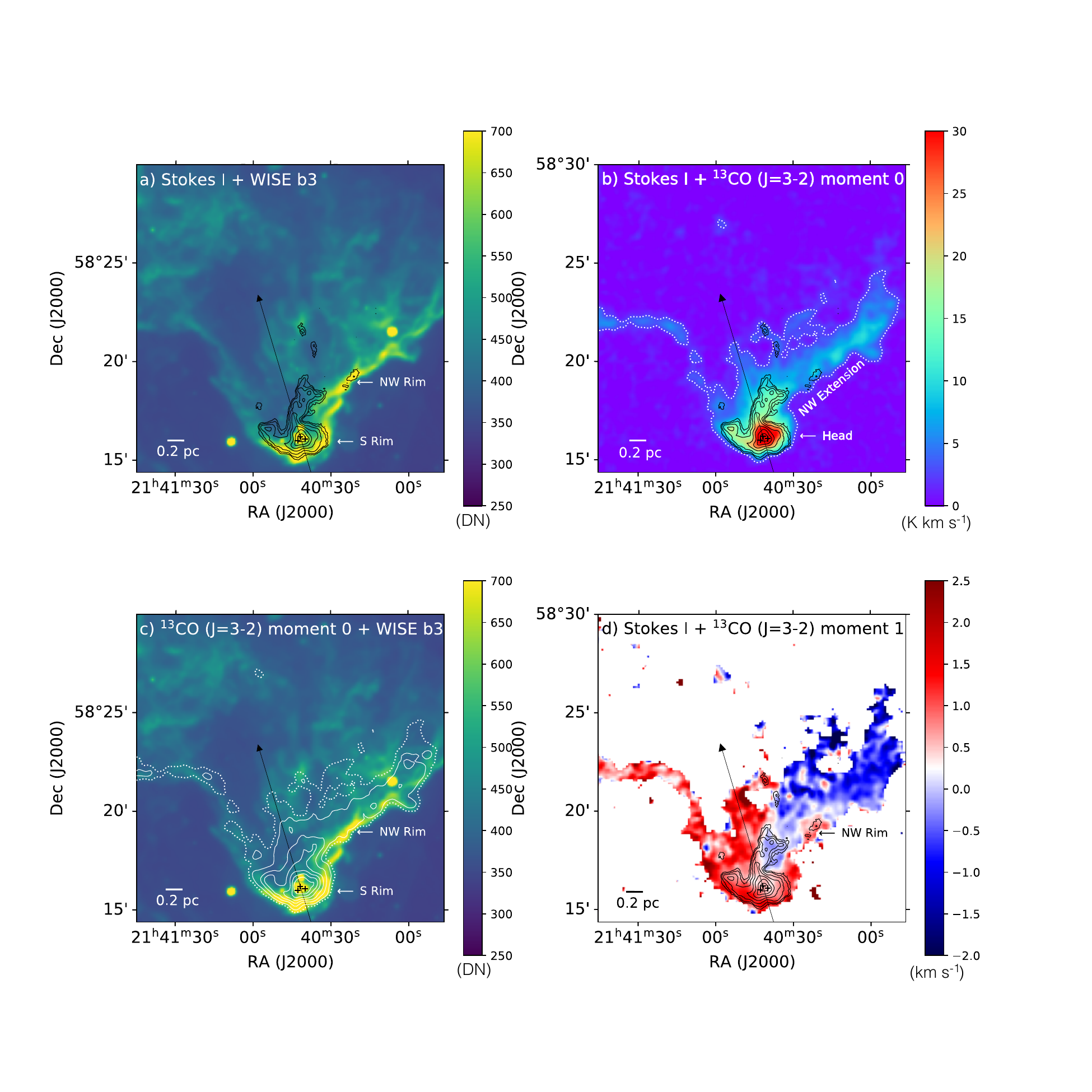}
 \end{center}
\caption{Comparison with the WISE band 3 and $^{13}$CO ($J=3-2$) moment 0/1 maps. 
The UV incident direction from HD~206267 by an arrow on each panel.
(a) The contours of the Stokes I intensity map is superimposed on the WISE band 3 image.
The contour levels at  a doubling scale-up rate are 5, 10, 20, 40, 80, 160, 320, and 640 ~MJy~sr$^{-1}$. 
The  flux density scale of the WISE band 3 is shown on the right side of the figure in image pixel units (DN).
(b) The contours of the Stokes $I$ intensity map are superposed on the $^{13}$CO ($J=3-2$) moment 0 map, of which the LSR velocity range is -5.0 to 6.0 km s$^{-1}$.  The contour of 2.5~K km s$^{-1}$ is shown by dotted lines.
(c) The contours of the moment 0 maps are superposed on the the WISE image, where the contours of  5, 10, 15, 20, 25, and 30~K km s$^{-1}$  are shown by solid lines and those of 2.5~K km s$^{-1}$ by dotted lines.  
(d) The contours of the Stokes I intensity map is superposed on the $^{13}$CO ($J=3-2$) moment 1 map.  
The velocity scale is shown on the right side in units of km s$^{-1}$.
}
 \label{fig3}
\end{figure*}

\begin{figure*}[ht!]
\begin{center}
\includegraphics[width=16cm]{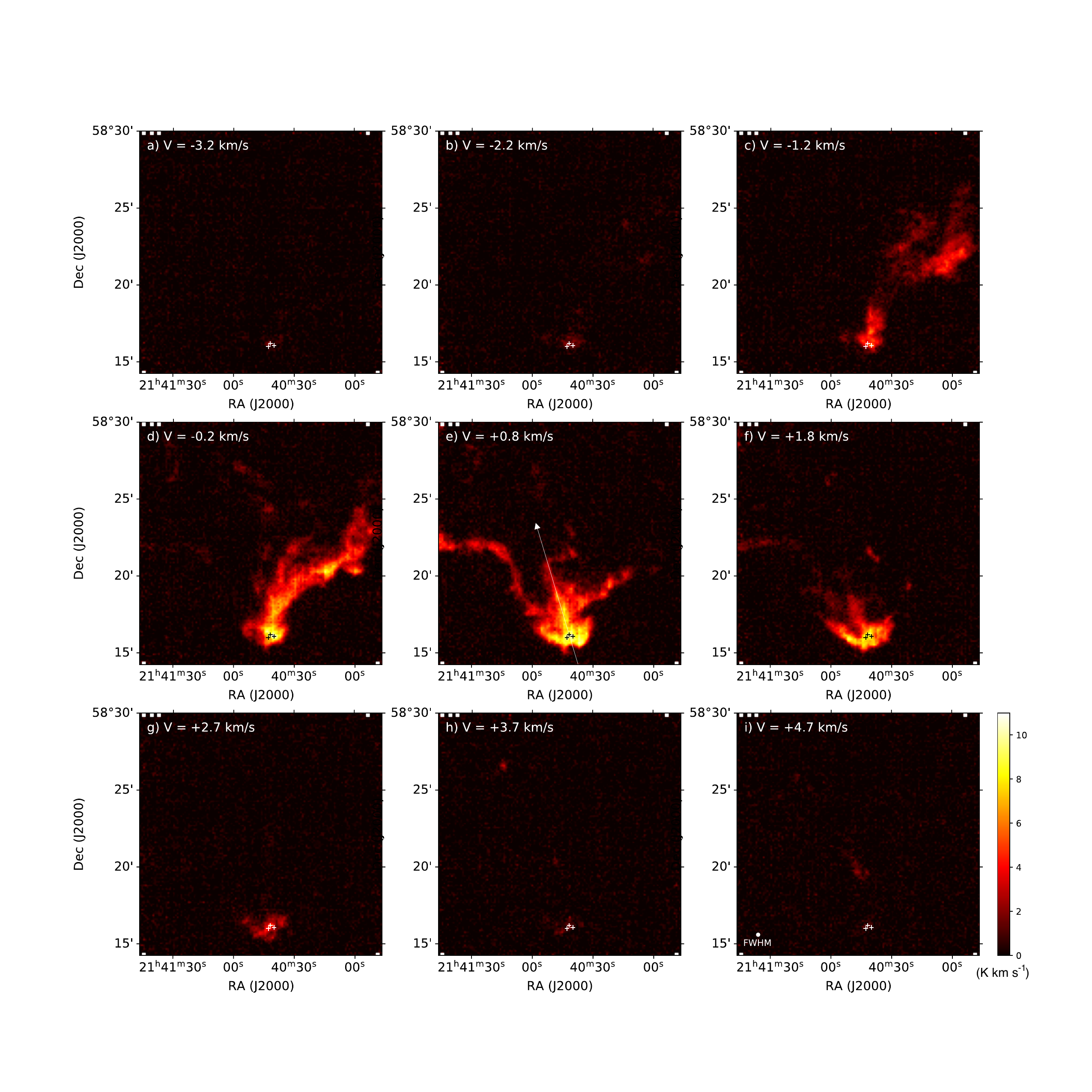}
\end{center}
\caption{Channel maps of $^{13}$CO ($J=3 - 2$).  
The central velocity is shown at the top left of each panel. 
The intensity scale is shown in K~km~s$^{-1}$ at the bottom right of the figure. 
The UV incident direction from HD~206267 is indicated by an arrow on panel e.
The positions of BIMA 1-3 are shown by white/black plus marks.
} 
\label{fig4}
\end{figure*}

The Stokes $I$ map shows a head-tail structure with an extension, i.e., a head with a tail (here, tail) and ear-like wings, and an extension part  to the northwest (here, the NW extension) that is very faint except for its root (figure \ref{fig2}b). 
This extension structure can be more clearly seen in $^{13}$CO ($J=3-2$), as shown in figure \ref{fig3}b.
The root of the NW Extension has a structure that may consist of a few cores (e.g., see \cite{2002aprm.conf..213S}).
The ear-like wing on the west side of the head (here, west wing)  is inconspicuous compared with the east side one (east wing), and these wing structures can be more clearly seen in the channel maps of the $^{13}$CO data as shown later in figure \ref{fig4}. 
A small concave surface can be seen toward the cloud tip.

In figures \ref{fig3}a, \ref{fig3}b, and \ref{fig3}d,  the Stokes $I$ contours (from $\sim 6 \sigma$ level of 5 MJy sr$^{-1}$ up to 640~MJy sr$^{-1}$) are superimposed on the images of  the WISE band 3, the $^{13}$CO moment 0 map, and the $^{13}$CO moment 1 map, respectively.
In addition,  the $^{13}$CO moment 0 map is superimposed on the WISE band 3 image (figure \ref{fig3}c).
Polycyclic aromatic hydrocarbon (PAH) molecules are considered to be tracers of photodissociation regions  (PDRs; \cite{1999RvMP...71..173H}), which are neutral regions of molecular cloud surfaces just behind the ionization fronts, indicating the ionization fronts laid on their immediate outside.
In the photodissociation regions, PAH molecules are excited by far-UV radiation that penetrates into the neutral surfaces, and give off fluorescent emission at 3.3, 6.2, 7.7, 8.6 and 11.3~$\mu$m bands (e.g., \cite{2002MNRAS.331...85R, 2003A&A...409..193U}). 
The WISE band 3 with a central wavelength of $\sim$11.6~$\mu$m would indicate the boundary between the neutral gas and the ionization front of the BRC, because its passband contains the 8.6 and 11.3~$\mu$m bands of PAH \citep{2010AJ....140.1868W}.
As shown in figure  \ref{fig3}a, the dense part of the cloud ($\gtrsim$5 MJy sr$^{-1}$ in the Stokes $I$ image) is enclosed by the bright emission in the 12~$\mu$m band emission (i.e., ionization front) on the exciting star side; here we refer this bright emission on the south side as the S Rim. 
The area with $^{13}$CO intensity of $\gtrsim$10~K~km~s$^{-1}$ in the moment 0 map is similar to the Stokes $I$ image in shape,  and is facing the bright 12~$\mu$m emission on the exciting star side. 
The less dense part of $\lesssim$10~K~km~s$^{-1}$, extending toward the north west of the head part, is also facing the bright 12~$\mu$m emission; we refer this emission as the NW rim (figures \ref{fig3}a and \ref{fig3}c).
In the NW extension, the $^{13}$CO intensity is higher on the exciting star side than the opposite side (figures  \ref{fig3}b and \ref{fig3}c), and the velocity gradient along the UV direction is seen, particularly from the brightest part of the NW rim toward the opposite side of the exciting star (figure \ref{fig3}d).
A narrow extension toward the north-east from the east wing is seen in $^{13}$CO (figures \ref{fig3}b and \ref{fig3}d).

In the optical, the whole appearance of IC~1396E/SFO~38 seems to be roughly symmetrical with respect to the UV incident direction (figure \ref{fig1}), and its structure appears to consist of a single object under the UV impact from the exciting star of IC~1396.
However, the Stokes $I$ image appears to be not symmetric to the UV direction (figure \ref{fig2}). 
While the tail of the head is parallel to the UV direction (NNE; PA$\sim 17\arcdeg$), its north-east end  gradually curved to the north-west direction (NW), i.e., not parallel to the UV direction.  
Also as mentioned above, the ear-like wings are not symmetrical in intensity.
In addition, the whole distribution of $^{13}$CO is much more asymmetrical with respect to the UV  direction.  
It is much stronger on the west side along the NW rim than on the east side (figures \ref{fig3}b and \ref{fig3}c).

To further examine the asymmetry and the velocity structure of the cloud, we made velocity channel maps of $^{13}$CO ($J=3-2$)  (figure \ref{fig4}).
In this figure, the velocity structure also shows asymmetry as a whole with respect to the UV direction, 
although the head-tail structure with ear-like wings, which is seen in the higher intensity area in panel e ($V_{\rm LSR} = 0.8$~km~s$^{-1}$)  and  in panel f ($V_{\rm LSR}$ = 1.8~km~s$^{-1}$), is fairly symmetry.
Such a head-tail structure could be expected in a radiatively imploded cloud, where the converging flow caused by the ionization front advances toward the symmetry axis of the spherical cloud (e.g., \cite{1994A&A...289..559L, 2009ApJ...692..382M, 2009A&A...497..649B, 2011ApJ...736..142B, 2014MNRAS.444.1221K, 2015MNRAS.450.1017K}), though the morphology of the imploded cloud would depend on its initial condition and evolutional stage. 
The concave structure of the cloud tip is clearly seen in the panel e ($V_{\rm LSR} = 0.8$~km~s$^{-1}$), and would be also expected for some initial condition and evolutional stage  (For more details on the concave structure formation, see subsection 3.2 of \cite{2010ApJ...717..658M}. See also Figure 12 of \cite{2009ApJ...692..382M}).
The velocity of this head-tail structure is redshifted relative to the ionized gas velocity by $\sim2$~km~s$^{-1}$ or more \citep{1980MNRAS.192..179P}, suggesting that the cloud is accelerated toward the far side of the H\emissiontype{II} region due to the UV impact. 
The narrow $^{13}$CO extension toward the north-east from the east wing has its velocity mostly at $\sim0.8$~km~s$^{-1}$ and faintly at 1.8~km~s$^{-1}$, indicating the UV impact from the exciting star.
The extended structure toward the NW from the head part (NW extension) has the velocity of mainly $\sim -1.7~{\rm to}~+0.2$~km~s$^{-1}$ that is slightly red-shifted relative to the ionized gas velocity (the panels c and d of figure \ref{fig4}). 
On the other hand, this velocity range is blue-shifted relative to that of the head region. There is a possibility that this velocity difference is due to accretion from the NW extension on to the head region and/or due to rotation of the NW extension along its elongation.  
Within the NW extension, the area with higher $^{13}$CO intensity on the side facing the ionization front (NW rim) has larger red-shifted velocity than the lower intensity side and this velocity gradient is along the UV direction (NNE-SSW) (figure \ref{fig3}d).  Considering these changes of the velocity and $^{13}$CO intensity within NW-Extension, NW extensions are most likely to be affected by UV, although the possibility of accretion/rotation cannot be completely ruled out.

\subsection{Magnetic Field Structure}\label{subsec:mf}

\begin{figure*}[ht!]
\begin{center}
\includegraphics[width=19cm]{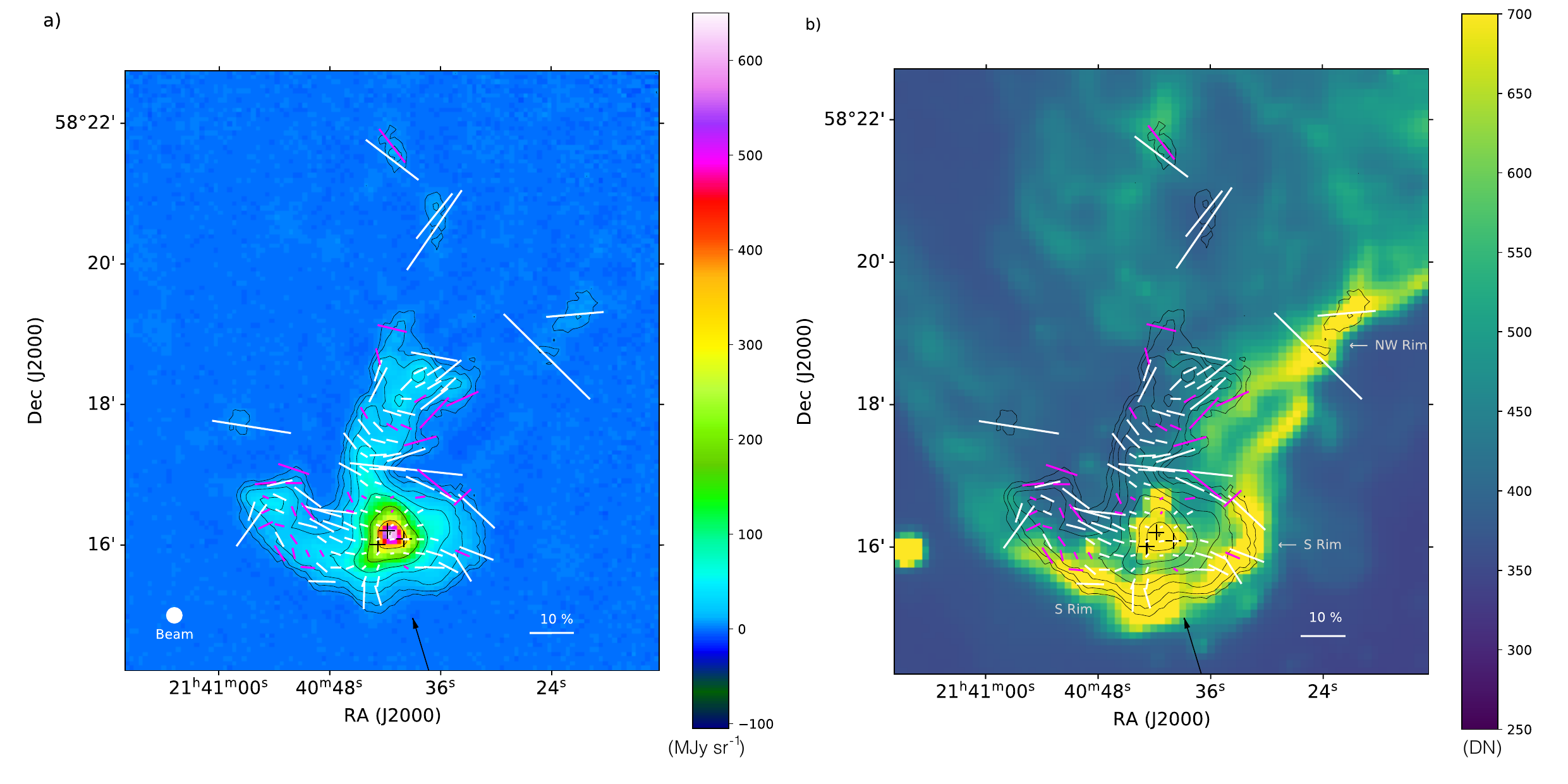}
\end{center}
\caption{Magnetic field orientation vectors superimposed on the Stokes $I$ and WISE band 3 images. 
The vectors show magnetic field orientations by rotating the measured polarization angles by 90$\arcdeg$. 
The vector lengths indicate the polarization degree $P$ and the legend of 10~\% polarization is shown at the bottom right of each panel. 
The polarization measurements with $P/\Delta P>3.0$  and $3.0 \geq P/\Delta P>2.0$ are shown by white and magenta vectors, respectively. 
The UV incident direction from HD 206267 is indicated by a black arrow on each panel.
(a) The Stokes $I$ image with the contours of  5, 10, 20, 40, 80, 160, 320, and 640~MJy~sr$^{-1}$.
(b) The WISE band 3 image with the contours of the Stokes I image.  
The scale level is shown in image pixel units (DN).
}
 \label{fig5}
\end{figure*}

\begin{figure*}[ht!]
\begin{center}
\includegraphics[width=15cm]{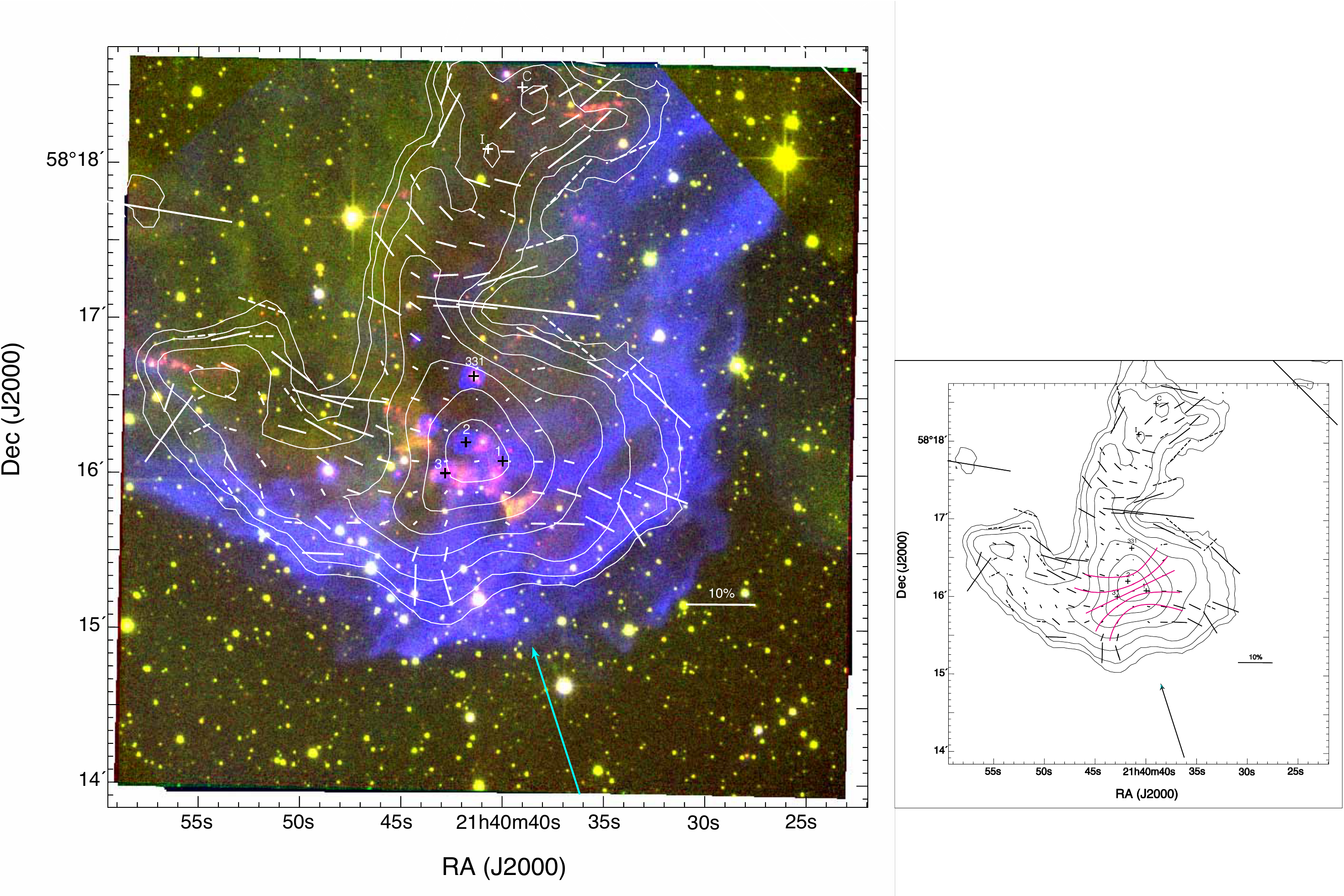}
\end{center}
\caption{Magnetic field orientation vectors superimposed on the pseudo-color image made from the image data at $K$s, $H$, and 8$\mu$m (Spitzer/IRAC channel 4), where red, $K$s; green, $H$; blue, 8$\mu$m (the left panel).
The passband of IRAC channel 4 contains the 7.7 and 8.6~$\mu$m PAH emissions \citep{2010AJ....140.1868W}.
The Stokes $I$ contours are also shown in the same way as in figure \ref{fig5}. 
The polarization measurements with $P/\Delta P>3.0$  and $3.0 \geq P/\Delta P>2.0$ are shown by solid and dashed-lines vectors, respectively. 
The three BIMA sources are shown by numbered plus marks.
Three more young stellar sources that are driving outflows (source C, source I, and  \#331, \cite{2001A&A...376..271C, 2009A&A...504...97B}), are also shown.
Note that the IRAC channel 4 data are absent in the upper right and left corners.
The field vectors around the peak intensity position of the Stokes $I$ appear to show an orderly pinched or hour-glass pattern (see the magenta lines on the right panel).
}
\label{fig6}

\end{figure*}

In figure \ref{fig5}, the polarization vectors rotated by 90$\arcdeg$, i.e., the directions of magnetic fields, are shown superimposed on enlarged images for the dense part of the cloud of the Stokes I and the WISE band 3, respectively. 
Here, we included  the data  with $3> P/\Delta P> 2$ and $\Delta P < 4 \%$ (see the criteria for inclusion in subsection 4.4 of \cite{2018ApJ...859....4K}) in addition to those with $P/\Delta P > 3$ .
A polarization map for a more limited area, superimposed on the pseudo-color image made from the $K$s and $H$ band data, and the Spitzer band 4 data, is also shown in figure \ref{fig6}. 
The Spitzer band 4 with a central wavelength of $\sim 7.9~\mu$m would indicate the boundary between the neutral gas and the ionization front of the BRC, because its passband contains the 7.7, and 8.6~$\mu$m bands of PAH \citep{2008PASP..120.1233H}.

\begin{figure*}[ht!]
\begin{center}
\includegraphics[width=15cm]{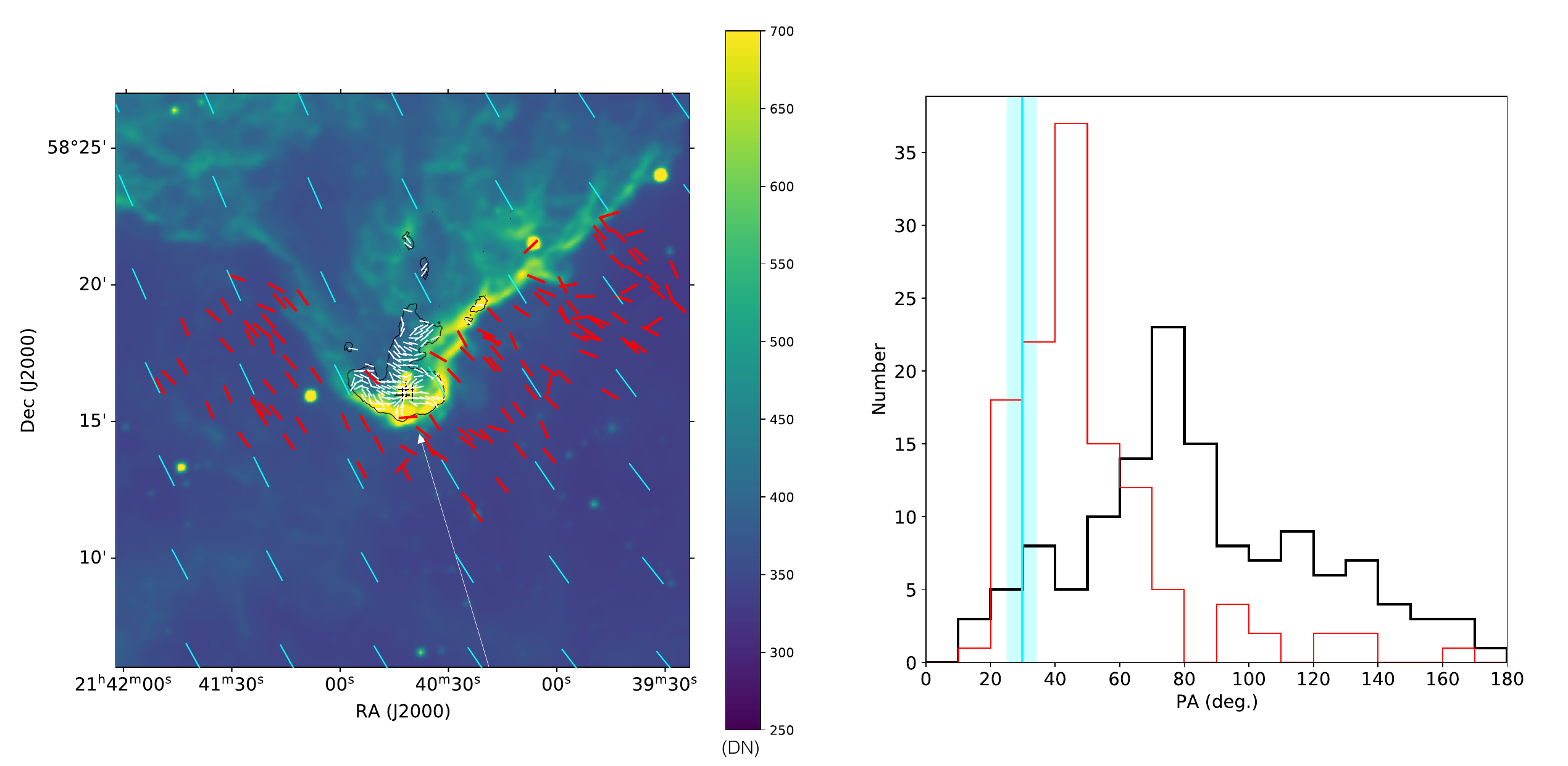}
\end{center}
\caption{Comparison among the polarization angles measured in this work (rotated by 90$\arcdeg$), those of the peripheral stars measured in the optical \citep{2018MNRAS.476.4782S}, and the global magnetic field around IC~1396E/SFO~38  from the Planck data.  
All the vectors of each data are of equal length (i.e., independent of the polarization degree) to make it easier to understand the direction change of the field. 
The left panel is the WISE band 3 image with the contour of 5~MJy sr$^{-1}$ of the Stokes $I$ intensity.
The polarization angles of the data selected in this work (see subsection 3.2) are shown by white vectors on the left image and by black lines on the right histogram. 
Those of the optical polarimetry are shown by red vectors ($P/\Delta P > 2$)  on the left image and by red lines on the right histogram.
The magnetic field directions of Planck are shown by cyan vectors on the left image, and their average and one sigma deviation by a cyan line and a light cyan band on the right histogram.
}
\label{fig7}
\end{figure*}

The optical polarization vectors of the peripheral stars \citep{2018MNRAS.476.4782S} and the field vectors from the Planck data are shown for comparison with our POL-2 polarization vectors rotated by 90$\arcdeg$ (figure \ref{fig7}).
The magnetic field orientation inside the cloud seems inconsistent with those of the outside from the optical and Planck polarimetry. 
The global field direction around the IC~1396E/SFO~38 cloud is estimated to be $29.\arcdeg8\pm4.\arcdeg6$ on average from the Planck data within the image of figure \ref{fig7}.
The angle distribution  in the optical (i.e., the ambient field direction) with a peak between 40$\arcdeg$ and 50$\arcdeg$ is slightly larger than that of Planck, but the field orientation appears to have a tendency to get closer to that of Planck as the data point moves away from the cloud, i.e., the field angle tends to slightly increase as the data point approaches the cloud.
The optical distribution is narrower than that of the POL-2 observation with a peak between 70$\arcdeg$ and 80$\arcdeg$. 
These differences of the magnetic fields would also suggest, at least in part, that the UV radiation may affect the magnetic fields of the cloud. 
A similar difference has also been seen in the IC~59 cloud, which is the nearest BRC in the vicinity of the B-type star Gamma Cas \citep{2017MNRAS.465..559S}. 

\begin{figure*}[ht!]
\begin{center}
\includegraphics[width=16cm]{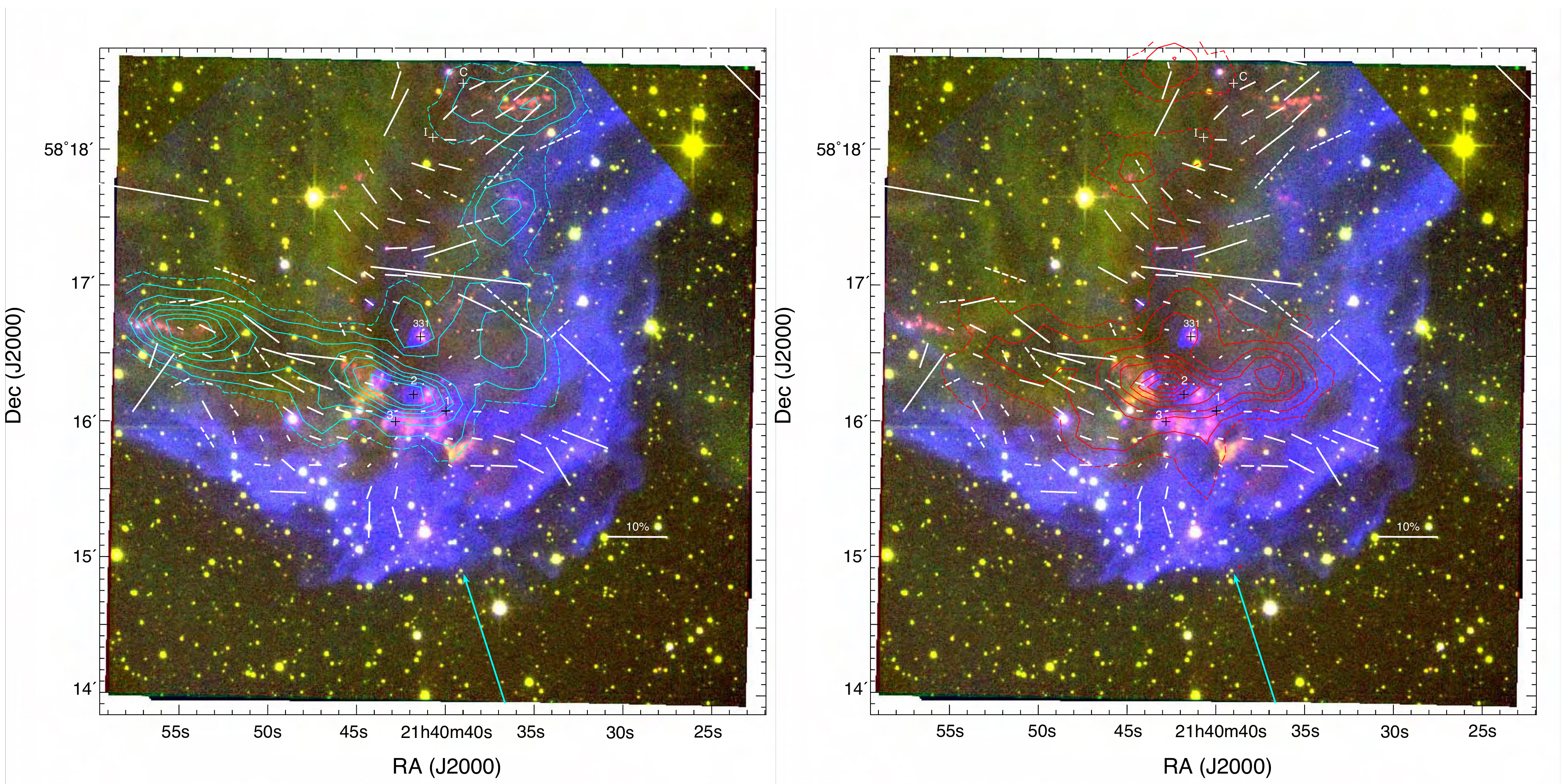}
\end{center}
\caption{Comparison of the magnetic field orientations with the CO ($J$=3-2) outflow lobes. 
The background image is same as that of figure \ref{fig6}. 
The integration ranges of the blue- and red-shifted emission are from -22.5 to -5.5 km s$^{-1}$ and from 5.5 to 22.5~km s$^{-1}$, respectively. 
The dashed lines are  the contours of 2.5~K km s$^{-1}$ ($\sim3 \sigma$),  and the solid lines are the contours of  every 10.0~K~km~s$^{-1}$ from  5.0~K~km~s$^{-1}$.
The UV incident direction from HD~206267 is indicated by a cyan arrow.
}
\label{fig8}
\end{figure*}

In the east wing that faces the S rim, the field vectors appear to be along this wing or the rim, except its easternmost point (figures  \ref{fig5} and \ref{fig6}).
The magnetic field direction along the east wing is PA $\sim50\arcdeg$ on average, similar to the polarization angles of the periphery stars, $\sim40\arcdeg$-~$50\arcdeg$, i.e., the ambient magnetic field. 
A little north of this east wing area, blue-shifted and red-shifted outflow lobes are extended from the central part of the head  (figure \ref{fig8}). 
In this lobe area, the  field vectors have a mean PA of $\sim70\arcdeg$, slightly different from that of East Wing, with a few times larger polarization degrees. 
This PA value of $\sim70\arcdeg$ is almost same as the directions of the outflow lobes. 

Around the head part where the three BIMA sources are located, the field vectors appear to show an orderly pinched or hour-glass pattern (figure \ref{fig6}). 
This field distortion is likely to be due the gravitational contraction of the head part, because the center of this pattern seems to be coincident with the peak of the Stokes $I$ intensity, i.e., the possible center of gravity of the head part. 
Note that the axis of the hour-glass pattern seems to be nearly perpendicular to the UV direction. 
Toward a small protrusion on the south side of the head part (see figures \ref{fig2}b and \ref{fig6}), the field vectors appear to run along this protrusion that faces a small concave structure at the head tip, which might have been created by the RDI mechanism (e.g., due to compression/erosion stronger at the tip than its surroundings; \cite{2009ApJ...692..382M, 2010ApJ...717..658M}).  
On the west side of the head part, there is an area where the polarization is not detected, though the Stokes $I$ intensity is rather strong there  (figure \ref{fig6}). 
In this undetected area with low polarization degrees, a red-shifted outflow lobe is seen and appears to be enclosed by the Spitzer 8~$\mu$m emission inside the S rim (figure \ref{fig8}). 
The non-detection is likely to be  due to either the field direction being almost in the line-of-sight or a chaotic field.
Just to the north and south sides of this undetected area, the field directions seem close to that of the ambient magnetic field.
The field vectors on this south side vertically crosses the S rim.
Between the tail stretched from the head part (tail) and the root of the NW extension, the field vectors also have PAs close to the ambient magnetic field.
On the other hand, in the more northern part of the root of the NW extension, which faces the NW rim, the field vectors seem to run parallel to the NW rim, i.e., perpendicular to the UV incident direction (figures \ref{fig5} and \ref{fig6}).

\section{Estimation of Magnetic Field Parameters}\label{sec:emf}

\begin{figure}[t]
\includegraphics[width=9cm]{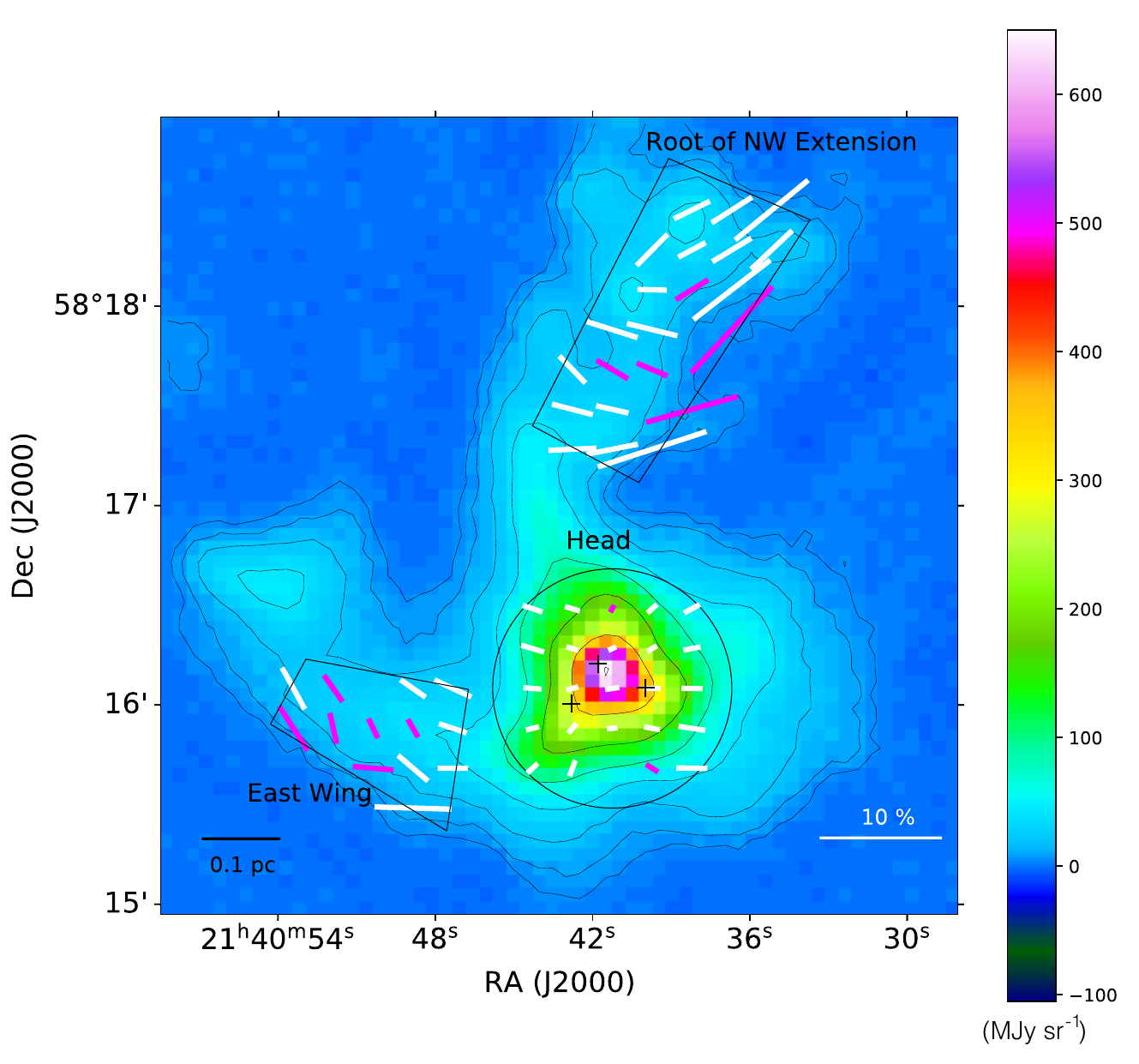}
\caption{Selected three areas to estimate the magnetic field strength on the Stokes $I$ map.
The field vectors that are used for the estimate are presented, enclosed by black lines.
The intensity color scale in MJy~sr$^{-1}$ is shown on the right side of the image.
The intensity contours and the polarization vectors are shown in the same way as in figure \ref{fig5}.
}
 \label{fig9}
\end{figure}

As shown in the previous section, the field direction differs in each area, but each field seems ordered in some degree.
The ordered field would imply an (at least) moderately strong field. 
So, we tried to roughly estimate the magnetic field strength and magnetic stability in three selected areas; the head part where the three BIMA sources are located (head), the east wing, and  the root of the NW extension, in order to examine the role of the magnetic field in terms of  cloud support (figure \ref{fig9}). 

The most commonly used way to estimate the magnetic field strength so far is the Davis-Chandrasekhar-Fermi (DCF) method \citep{1951PhRv...81..890D, 1953ApJ...118..113C} using the polarization data. 
With the DCF method, the strength of the plane-of-sky (POS) component of the magnetic field can be estimated as   

\begin{equation}
  B_{\rm pos} = f\sqrt{4\pi \rho}\frac{\sigma_{v}}{\sigma_\theta}, 
\end{equation}  
where $\rho$ is the mean volume density of the cloud, $\sigma_v$ is the line of sight velocity dispersion, 
and $\sigma_\theta$ is the angular dispersion of the polarization vectors, and $f$ is a correction factor. 
As a correction factor,  $f\sim0.5$ has been often used for $\sigma_\theta \lesssim 25\arcdeg$ \citep{2001ApJ...546..980O} .

As shown in figure \ref{fig9},  the field vectors are not necessarily aligned straight, but curved to some degree.
In the curved case, we would overestimate the angular dispersion when we simply calculate it. 
Then, it is necessary to remove the contribution from the curved field by assuming the field shape obtained by fitting the curved field data or by using the angular dispersion function method  (ADF method; \cite{2009ApJ...696..567H,2009ApJ...706.1504H}).   
For the angular dispersion estimate here, we used the ADF method, which has no need for assuming the field shape.
While the method of \citet{2009ApJ...706.1504H} is the improved version of \citet{2009ApJ...696..567H}, which can estimate the turbulent correlation length as well as the angular dispersion, it needs more data points than that of \citet{2009ApJ...696..567H}. 
Unfortunately, the number of data points in each area seems to be not enough to apply the method of \citet{2009ApJ...706.1504H}.
Therefore, we used that of \citet{2009ApJ...696..567H}.
The averaged square of the differences in $\theta$ (PA) between the $N(l)$ pairs of vectors separated by displacements $l$, 
i.e., $\theta(l) \equiv \theta ({\bf X}) - \theta({\bf X}+{\bf l})$, can be calculated as 
\begin{equation}
  <\Delta\theta(l)^2>=\frac{1}{N(l)}\Sigma_{i=1}^{N(l)}[\theta ({\bf X}) - \theta({\bf X}+{\bf l})]^2,  
\end{equation}
where $< > $ denotes an average, ${\bf X}$ denotes the position of the polarization vector, and $l=|{\bf l}|$. 
In the case that the displacement $l$ exceeds the correlation length $\delta$,  
which characterizes a turbulent component $B_{\rm t}$, and is smaller than the scale $d$, 
which is the typical length scale for variations in the curved magnetic field  $B_0$, 
i.e., if $\delta < l < d$, the square of the total measured dispersion function is expected to become 
\begin{equation}
  <\Delta\theta(l)^2> _{\rm tot} \simeq b^2 + m^2 l^2+\sigma_{\rm M}^2(l),  
\end{equation}
where $m$ is a coefficient that indicates the contribution from $B_0$, $\sigma_{\rm M}(l)$ 
is a measurement uncertainty,  and $b^2$ is the intercept of the squared function at $l=0$.
We have applied this formula to the three selected areas and the fitting results are shown in the Appendix.
Here, $b$ is linked to $B_{\rm t}$ and $B_0$ as 

\begin{equation}
  \frac{<B_{\rm t}^2>^{1/2}}{B_0} =\frac{b}{\sqrt{2-b^2}}, 
\end{equation}
where the total magnetic field $\bf{B}$ consists of deterministic ($\bf{B}_0$) and turbulent ($\bf{B}_{\rm{t}}$) components. 
We derived the dispersion angle $\sigma_\theta$ as $\sigma_\theta=< B_{\rm t}^2>^{1/2}/B_0$ \citep{2009ApJ...696..567H}.

The velocity dispersion, $\sigma_v$, was derived from the C$^{18}$O ($J=3-2$) data of the JCMT archive, since the distribution of C$^{18}$O ($J=3-2$)  seems similar to that of the Stokes I. 
However, the intensity of C$^{18}$O ($J=3-2$) is fairly weak compared with that of $^{13}$CO ($J=3-2$).
So, we made the averaged profiles of C$^{18}$O ($J=3-2$) of the three selected areas and obtained the velocity dispersion by Gaussian fitting.
To examine the effect of velocity gradient within each area, we also obtained velocity dispersions of every data points within each selected area by using the convolved data with a Gaussian kernel (stddev = 1; $\sim$0.056~km~s$^{-1}$) and averaged them.   
As a result, we found that the values of both method are almost the same, so we used those obtained with the former method. 
To obtain the mean column density of molecular hydrogen from the Stokes I flux, we adopted the dust temperatures of $T_{\rm d} = 23$~K and 18~K for the head part and the root of the NW extension, respectively, by referring the temperatures derived from the NH$_3$ observation of SFO~38 (figure 3 of \cite{1993ApJ...404..247S}). 
For the Eeast wing, no temperature data are available, but this area can be expected to have somewhat higher temperature because of its direct contact to the ionization front (S rim), like the temperature of the head tip of $\gtrsim25$~K in the NH$_3$ observation.
So we adopted  $T_{\rm d} = 25$~K for the east wing.
The column density of each area was derived as follows:
\begin{equation}
N ({\rm H_2}) = \frac{I_\nu}{B_\nu(T_{\rm d})\kappa_\nu\mu m_{\rm H}}, 
\end{equation}
where $I_\nu$ is the observed surface brightness at frequency $\nu$ , $B_\nu(T_{\rm d})$ is the Planck function, $\kappa_\nu$ is the dust opacity per unit (dust+gas) mass, $\mu$ is the mean molecular weight of 2.8, and $m_{\rm H} $ is the hydrogen mass.
We adopted $\kappa_\nu = 0.1(\nu/1000{\rm GHz})^\beta$~cm$^2$~g$^{-1}$ and the dust emissivity index of $\beta$ = 2, as in \citet{2010A&A...518L.106K}.
The averaged column density of each area was obtained and the volume number density was derived by dividing it by the width/diameter of each area (table \ref{tab:bparms}). 

As mentioned above, $f\sim0.5$ has been often used for $\sigma_\theta \lesssim 25\arcdeg$ as a correction factor of the DCF method  \citep{2001ApJ...546..980O}.
\citet{2009ApJ...706.1504H} showed an underestimate of the measured angular dispersion due to the signal integration through the thickness of the cloud and the area of the telescope beam. 
So, they suggested that a correction factor is $f\sim 1/\sqrt{N}$, where $N$ is the number of the independent turbulent cells contained within the gas/dust column probed by the telescope beam. 
They defined $N$ as follows: 
\begin{equation}
N = \frac{(\delta^2+2W^2)\Delta^\prime}{\sqrt{2\pi}\delta^3}, 
\end{equation}
where $W$ is the telescope beam radius, $\delta$ is the turbulence correlation length, and  $\Delta^\prime$ is the effective depth of the cloud.
The telescope beam radius is $\sim7\arcsec$ (0.031 pc at 900 pc).
To evaluate the correction factor, we need to know the correlation length and the effective depth of the cloud, but these are unknown for the IC~1396E/SFO~38 cloud.
So, we adopted plausible values for these unknowns as follows. 
\citet{2022FrASS...9.3556L} showed that the correlation lengths from the literature with the ADF method have a wide range of values (see their figure 2), and suggested a correlation trend between the turbulence correlation length and the telescope spatial resolution. 
They also suggested that the cloud scale/density may affect the local turbulence. 
We searched for observations with a spatial resolution similar to ours for regions with densities similar to the IC~1396E/SFO~38 cloud. 
In OMC-1HII (see figure 8 of \cite{2019ApJ...872..187C}), which has similar scale and column density, the turbulence correlation length of 19~mpc was estimated at 214~$\mu$m (HAWC+) with spatial resolution that is slightly better than ours (see both of table 1 of \cite{2022ApJ...925...30L} and table 4 of \cite{2019ApJ...872..187C}).
Also in OMC-1Bar, which has similar scale and slightly large column density, as well as in  OMC-1HII,  the values of 15-19~mpc were estimated at shorter wavelengths (with higher resolution), except at 53~$\mu$m in OMC-1HII (7.4~mpc). 
Although the correlation trend was suggested above, most of the analysis results appear to show that the estimated correlation lengths are distributed around about 20~mpc in figure 2 of \citet{2022FrASS...9.3556L}. 
Here, we adopted $\delta = 20$~mpc as the turbulence correlation length.
A method, using the normalized autocorrelation function of integrated polarized flux, that approximately derives the effective cloud depth was introduced (see equation (51) and figure 1 of \cite{2009ApJ...706.1504H}). 
We first tried to apply this method for the head part and obtained a distance ($\lesssim 0.1$ pc) where the autocorrelation function becomes half.
If we adopt this value or twice as the cloud depth, it would be $\sim 0.1-0.2$ pc.
Unfortunately, this method is not valid for the other two areas, due to the asymmetry structure and/or insufficient data points.
The widths of the three areas shown in figure \ref{fig9} are $\sim0.14-0.31$~pc with an average of $\sim0.21$~pc and the widths of the half intensity are $\sim$0.1 pc in all three areas. 
So, we adopt an effective depth of $\sim0.1-0.2$~pc and obtain the correction factor of $\sim0.2-0.3$. 
Here, we estimated the magnetic strength by adopting the correction factor of 0.25 (table \ref{tab:bparms}). 

We also evaluate the magnetic stability of each area with the estimated  field strength. 
The mass-to-flux ratio normalized by the critical mass-to-flux ratio for the magnetic stability is calculated as follows: 
\begin{equation}
\lambda_{\rm obs} = \frac{\mu m_{\rm H}N({\rm H}_2)}{\lambda_{\rm crit}B_{\rm pos}}.
\end{equation}
The critical ratio for the magnetic stability was given by \citet{1978PASJ...30..671N}, $\lambda_{\rm crit}=1/\sqrt{4\pi^2G}$,  where G is the gravitational constant. 
The evaluated normalized ratios are also presented in table \ref{tab:bparms}. 

The Alfv\'{e}n Mach number was also calculated for each area as follows:  
\begin{equation}
M_{\rm A} = \frac{\sigma_v}{V_{\rm A}}=\sqrt{4\pi \rho}\frac{\sigma_v}{B_{\rm pos}}.
\end{equation}
This number is a criterion for whether the cloud is super-Alfv\'{e}nic ($M_{\rm A}>1$) or sub-Alfv\'{e}nic ($M_{\rm A}<1$) .
Although the calculated values indicate sub-Alfv\'{e}nic (table \ref{tab:bparms}), they would become $M_{\rm A}>1$  by taking into account the 3D correction factor for the velocity dispersion of $\sqrt{3}$.   
On the other hand, these numbers would decrease by taking the 3D field structure into account.

\begin{table*}
  \tbl{Approximate sizes and averaged physical/magnetic parameters for selected areas. }{%
  \begin{tabular}{lccccrcc}
      \hline
      Area & $\overline{N}$(H$_2$) & $\overline{n}$(H$_2$) & $\sigma_v$  & $\sigma_\theta$ & $B_{\rm pos}$ & $\lambda_{\rm obs}$ & $M_{\rm A}$\\ 
      Name & ($\times10^{21}$ cm$^{-2}$) & $\times10^{4}$ cm$^{-3}$) & (km s$^{-1}$) & (deg.) & ($\mu$G) & &\\
      \hline
      Head & 47.2 & 4.9 & 0.93$\pm$0.02 & 19$\pm$4 & 120$\pm$27 & 3.0$\pm$0.7 & 0.8  \\
      Root of  NW Ext.& 9.9 & 1.8 & 0.54$\pm$0.03 & 11$\pm$3 &  70$\pm$20 & 1.1$\pm$0.3 & 0.6\\
      East Wing           & 7.0 & 1.6 & 0.47$\pm$0.04 & 19$\pm$1 & 32$\pm$3 & 1.6$\pm$0.2 & 0.8\\
      \hline
    \end{tabular}}\label{tab:bparms}
\end{table*}

\section{Discussions} \label{sec:discussions}

\subsection{Magnetic Field} 
 
The magnetic field strengths were estimated in the three areas (table \ref{tab:bparms}), although these estimates might have certain ambiguities due to the telescope resolution, which is not always high enough to resolve the cloud structures, i.e., a small number of the field vectors. 
In the head,  our rough estimates indicate field strength of $B_{\rm POS}\sim120~\mu$G and a  mass-to-flux ratio of $\lambda_{\rm obs}\sim3.0$ that is normalized by the critical mass-to-flux ratio  \citep{1978PASJ...30..671N}.
In the root of the NW extension, $B_{\rm POS}\sim70~\mu$G and $\lambda_{\rm obs}\sim1.1$ are calculated.
In the east wing,  $B_{\rm POS}\sim30~\mu$G and $\lambda_{\rm obs}\sim1.6$ are estimated.
These results do not take into account the magnetic field of the line-of-sight component, but may show a tendency that the head is more likely to be magnetically supercritical than the others, thus indicating the head area is most  likely to form stars.
In fact, signs of active star formation have been detected in the head. 
In the root of the NW extension, star formation signs have been reported (e.g., \cite{2012A&A...542L..26B}), but the mass-to-flux ratio is not as high as in Head. 
This lower value may be due to the averaging effect that is cased by its internal structure (e.g., a few cores and their low-density envelopes), as seen in the Stokes $I$ image of this area. 
The mass-to-flux ratio of the east wing is higher than that of the root of the NW extension.
Within our selected area of the east wing, there is only one Class~I/II candidate (MIR-80), where Class~I/II sources are defined as those that have Spitzer color ranges of [5.8]-[8.0]$\gtrsim$1.1 and [3.6]-[4.5]$\lesssim$0.7 by \citet{2010ApJ...717.1067C}. 
If this source is really associated with the east wing, the higher value of the mass-to-flux ratio does not contradict it, but if not, the higher value may indicate future star formation. 
The field strength of the east wing of $B_{\rm POS}\sim30~\mu$G is not so high compared with that of the Galactic regular field ($B\lesssim10~\mu$G; e.g., sub-subsection 7.3.2 of \cite{2001cre..book.....G}) and indicate that the ambient field strength seems to be less than $\sim30~\mu$G.  
Given that the IC~1396E/SFO~38 cloud is located in a star formation site, the ambient field strength is expected to be somewhat higher than that of the regular field value. 
If the ambient field strength is $\sim 10-20~\mu$G,  the field of the east wing might not have been so intensified by the UV impact. 

\citet{2011MNRAS.412.2079M} have conducted 3D MHD simulations to investigate the effects of initially uniform magnetic fields on the formation and evolution of elongated globules at the boundaries of H\emissiontype{II} regions. 
They found that a field initially perpendicular to the incident UV light is swept into to align with the cloud elongation direction during cloud evolution for the cases of weak and medium field strengths.
On the other hand, they found that a strong field would remain it original direction.
These results indicate that the weak, initial field direction is altered by the UV light during cloud evolution. 
Given the difference between the internal and ambient magnetic fields of the IC~1396E/SF~O38 cloud, a similar change in the field direction due to the UV impact can be expected although the physical conditions may not be the same as in their simulations.
In addition, the cloud compression due to the UV impact might enhance the field strength. 

\citet{2013ApJ...766...50M} conducted numerical modeling of a photoevaporative spherical cloud in two cases where the initial direction of the magnetic field is perpendicular or parallel to the UV incident direction, taking into account the magnetic pressure and heating by the UV from the exciting  star. 
In this modeling, they approximately evaluated field strength $B$ in a quasi-stationary equilibrium state after compression by the UV impact as follows: 
\begin{equation}
  B=B_0\bigr(\frac{\rho}{\rho_0}\bigr)^\alpha
\end{equation}
where $\rho$ is the density after compression, and $B_0$ and $\rho_0$ are the magnetic field strength and density of the initial state before compression. 
For the case of the initial field perpendicular to the UV direction, they mentioned that $\alpha$ is close to 1 at around the cloud tip with high density after reaching a quasi-stationary equilibrium state, while $\alpha$ is close to 0 at around the cloud tail with low density. This means that the field strength of the tip is hight due to the field remain in perpendicular to the UV direction and that the field strength of the tail (wings) is low due to the field parallel to the UV direction. 	
The weak field strength of the east wing can be explained by a similar process. 

\subsection{Cloud Structure}

As described in section \ref{subsec:cs}, the IC~1396E/SFO~38 cloud is not symmetric about the UV light from the exciting star and appears to consist of two parts on first glance, i.e., a head part with wings and a NW extension,  from the SCUBA-2 and $^{13}$CO ($J=3-2$) data.
Here, we wonder whether these two parts are separate (independent) or continuous (dependent)  on site.  
In the following, we try to consider two cases. 

\subsubsection{Separate Case}
First we consider the case that these two parts are separated and not continuous.  
In this case, they could be partly overlapped between the head-tail and the root of the NW extension just on the line of sight.   
The UV light from the exciting star would reach each surface separately, forming ionization fronts and photodissociation regions on each surface on the exciting star side.  
The head part is enclosed by the S rim on the exciting star side, but the root of the NW extension seems not to entirely border the NW rim, only its northern part of the root of the NW extension (figures \ref{fig3}a, \ref{fig5}b, and \ref{fig6}).  
If the UV light toward the southern part of the root of the NW extension is blocked by the head part, the southern part would be just behind the head part when viewed from the exciting star. 
However, such a coincidental situation may not be easily expected and no signs of their separation is evident from the 850~$\mu$m continuum emission that traces higher density. 
The separation cannot be seen from the WISE data either.
In addition, the $^{13}$CO line-of-sight velocity appears to vary from the head part to the root of the NW extension smoothly, not abruptly and the difference in velocities is also small. 
The only thing is that the head is redshifted and the NW extension is blue shifted (figures \ref{fig3}d).

\subsubsection{Continuous Case}

Secondly, we consider the case that these two parts are continuous and not separated. 
In this continuous case, the two-part-like structure might have been created from a single cloud. 
As is widely known, molecular clouds are generally filamentary and those located in the peripheries of H\emissiontype{II} regions  are generally expected to be illuminated by UV light from OB stars and have bright rims on the surfaces on the OB star sides. 
It would be expected that the IC~1396E/SFO~38 cloud is such a case.  
So, we consider the possibility that this two-part-like structure have been made from a single elongated cloud by referring the radiation-hydrodynamic simulations of prolate clouds at H\emissiontype{II} boundaries \citep{2015MNRAS.450.1017K}. 
They conducted radiation-hydrostatic simulations to investigate how uniform density prolate clouds evolve when illuminated by UV radiation from different incident directions. 
They succeeded to reproduce a variety of morphological structures of BRCs. 
Since some of their simulated BRCs appear to resemble the IC~1396E/SFO~38 cloud in morphology, we try to closely examine our results by referring to their simulations. 

\citet{2015MNRAS.450.1017K} defined the ratio of the physical ionization penetration depth to the characteristic depth of the cloud as $d_{\rm EUV}$ (equation (8) of \cite{2015MNRAS.450.1017K}) and expected that the prolate clouds are in the RDI-triggered shock dominant regions when $d_{\rm EUV}<<1$.
For the IC~1396E/SFO~38 cloud,  an ionizing flux ($F_{\rm EUV}$)  of $1.33\times 10^9$~cm$^{-2}$~s$^{-1}$ was predicted by the equation of the ionizing photon flux, while $F_{\rm EUV}=0.28\times 10^9$~cm$^{-2}$~s$^{-1}$ was measured based on the NVSS snapshot observation \citep{2004A&A...426..535M}.  
This difference may be attributed to the filter-out effect of the interferometer, and so $F_{\rm EUV}\sim1\times 10^9$~cm$^{-2}$~s$^{-1}$ may be appropriate.
Using this ionizing flux and other possible initial parameters (e.g., a number density of a few $\times10^3$ cm$^{-3}$, a major-to-minor axial ratio $\gamma$ of less than five,  an incident angle of $\sim 30\arcdeg-~60\arcdeg$), $d_{\rm EUV}$  of $<<1$ can be expected, suggesting that  the IC~1396E/SFO~38 cloud is in the RDI-triggered shock dominant region. 

\citet{2015MNRAS.450.1017K}  conducted three series sets of SPH simulations for prolate clouds with different initial parameters; geometry shapes ($\gamma$: the ratio of the semiminor axis to the semimajor axis.), UV inclination angles ($\varphi$), initial densities ($n$), and ionizing fluxes, in order to  examine the variations of the cloud evolution. 
For all the simulations, the initial cloud temperature of 60~K and  an incident UV phot flux at the cloud, $F_0=1.0\times 10^9$~cm$^{-2}$~s$^{-1}$  (or $2.0\times10^9$~cm$^{-2}$~s$^{-1}$) were adopted.  
They found that many asymmetrical BRCs, filamentary structures, and irregular horse-head structures could be developed with various initial conditions at the peripheries of H\emissiontype{II} regions.  
The simulations with initial higher density (G1200 series) revealed the formation of a single high-density core at the prolate cloud edge on the star-facing side of the cloud, except for the case where the incident angle of UV is parallel to the minor axis of the cloud, (e.g., figures 3, 4, and 6 of \cite{2015MNRAS.450.1017K}). 
In the parallel case (the incident angle is perpendicular to the cloud elongation), two high-density cores were formed at both cloud edges. 
In some particular case where the incident angle of UV is oblique (45$\arcdeg$, 60$\arcdeg$, and 75$\arcdeg$) to the minor axis of the cloud as well as the initial density is higher ($700-1200$~cm$^{-3}$) and the initial $\gamma$ is $2-3$, a nose structure with a high-density core appear at the tip of a higher-density filamentary region that is an elongated compression layer on the star-facing side (figures 6 and 10 of \cite{2015MNRAS.450.1017K}). 
The shocked velocity direction of the compressed gas due to the rocket effect by the evaporating gas would differ in the nose and filamentary regions. 
The velocity pattern of the nose seems to resemble to the convergence velocity pattern of RDI for the spherical cloud, while the velocity direction of the filamentary region is  perpendicular to the filament elongation/edge (e.g., figure 18 of \cite{2015MNRAS.450.1017K}).  
Although their figure 18 is for the low density case, but the velocity structure difference would be similar even in the higher density case by considering the similar "nose + filament" structure. 

Looking at the simulated "nose + filament" structure, the IC~1396E/SFO~38 appears to resemble to their oblique cases in morphology.
The head of IC~1396E/SFO~38 may correspond to the nose structure and the NW extension to the higher-density filamentary region, although the initial conditions might not be the same as those adopted in the simulations. 
\citet{2018MNRAS.476.4782S} already have shown another example (SFO~39) of the oblique case in the eastern periphery of IC~1396.
They mentioned that the SFO~39 cloud might have formed from an initial inclined prolate cloud, i.e., a cloud obliquely illuminated  by the exciting star of IC~1396, under the RDI effect. 
They found that the magnetic field direction on the star facing side of the cloud is different from the global field, but not different on the other side. 
They suggested that this change was due to the compression on the star facing side during the RDI process.

\begin{figure*}
\begin{center}
\includegraphics[width=12cm]{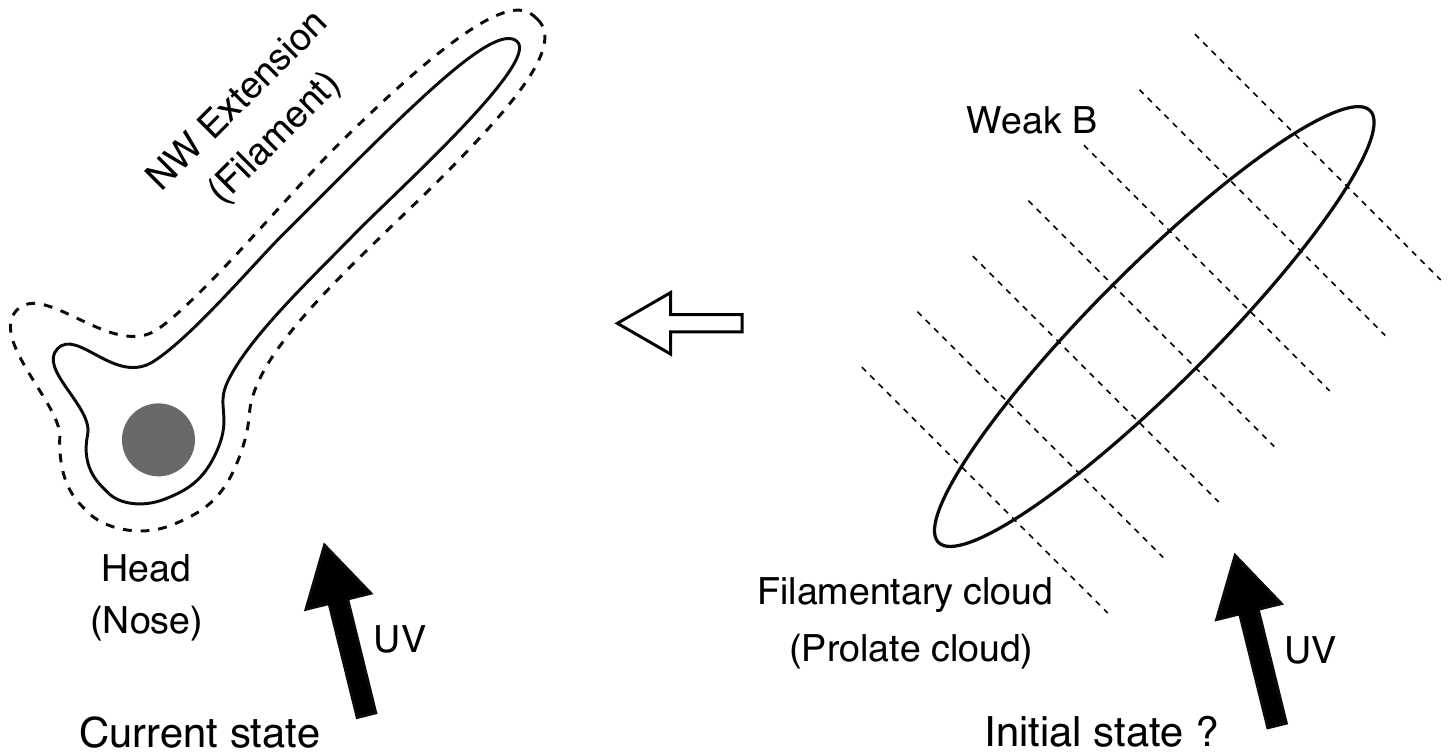}
\end{center}
\caption{Schematic drawing of the current and potential initial states of the IC~1396E/SFO~38 cloud. The correspondence to \citet{2015MNRAS.450.1017K} is noted in parentheses.
}
 \label{fig10}
\end{figure*}

\citet{2015MNRAS.450.1017K} did not take into account the magnetic field and stopped their calculations just before star formation occurred; i.e., magnetic effects and feedback from formed stars were not considered in their simulations. 
In the early evolution stage of UV-illuminated clouds with weak magnetic fields, the UV impact may be much more dominant than the magnetic effects and the cloud evolution may proceed in a similar way to non-magnetized clouds, given the weak field simulations of \citet{2011MNRAS.412.2079M}. 
If so, a similar "nose + filament" structure could form and then shift toward a late quasi-stationary equilibrium state where the magnetic field is important \citep{1990ApJ...354..529B, 2011MNRAS.412.2079M}.
In the IC~1396E/SFO 38 cloud, star formation already has occurred and the magnetic field with moderate strength is detected, indicating that the cloud situations differ from their simulations.
However, the weak initial magnetic field may not have had much effect on the early evolution of the  cloud, and  the field might have been strengthened during the UV interaction. 
Furthermore, the feedback from the formed stars appears to confined to part of the cloud and might not change the overall cloud morphology in large scale.
Thus, the fundamental physical processes could be expected to be similar to those of \citet{2015MNRAS.450.1017K} and we speculate that the two-part-like structure has been created from a single filamentary cloud obliquely illuminated by UV light (see figure \ref{fig10}).  
However, full MHD simulations are required for more detailed studies.

\section{Summary} \label{sec:summary}

We carried out 850 $\mu$m polarimetric observation for the bright-rimmed cloud IC~1396E/SFO~38 with SCUBA-2/POL-2. 
In addition to these POL-2 data, we used  the JCMT archival data of $^{12}$CO, $^{13}$CO, and C$^{18}$O ($J=3-2$), the Spitzer 8~$\mu$m and WISE 22~$\mu$m image data, and the $H$- and $K$s-band data to study  the UV impact on the magnetic field as well as the density and velocity structures of the cloud.  
The main results are summarized as follows:

\begin{enumerate}
  \item The Stokes $I$ map shows that the bright-rimmed cloud appears to consist of two parts;  a dense head part with a tail and ear-like wings, and a part extending to the north-west that is very faint except for its root, while it appears as one single cloud in the optical.
 The $^{13}$CO channel maps more clearly show the two-part-like structure, including the faint area of the north-west extension. 
 The dense head part and its wings are enclosed by an ionization front on the exciting star side, and the northwest extension is also facing an ionization front in parallel.  
 The whole cloud structure appears to be asymmetry to the UV incident direction, while the head part appears to have a structure nearly symmetry to the UV direction and to be similar to the structure of the radiation-driven implosion, expected for  a spherical cloud.
These would indicate that the initial shape of the clouds is not close to spherical.
  
  \item The polarization vectors of POL-2, rotated by 90$\arcdeg$, show that the angle distribution of the magnetic field inside the cloud with a peak of  $\sim 70\arcdeg-~80\arcdeg$ is wider than that of the outside with a peak of $\sim 40\arcdeg-~50\arcdeg$ (i.e., an ambient field direction). 
 Given the outside field angle measured in the optical and with Planck might be the initial field of the cloud, this difference/change in the angle distribution may be due to the UV impact on the magnetic field of the cloud during its evolution. 
 The change in the field angle distribution within and outside the cloud might suggest that the initial field was weak and have been altered in direction during the cloud evolution, by referring to the previous 3D MHD simulations of globules at the boundaries of H\emissiontype{II} regions \citep{2011MNRAS.412.2079M}. 
In the head part where three BIMA continuum sources are located, an orderly hour-glass pattern of the field centered at the peak of the Stokes I intensity with its axis perpendicular to the UV incident direction is seen, suggesting the gravitational contraction of the head part. 
 In the east wing extended from the head part and the root of the northwest extension, the fields along the ionization fronts are seen. 

  \item  We estimated the field strength and magnetic stability of three areas within the cloud, using the Davis-Chandrasekhar-Fermi method. 
  The average plane-of-sky field strengths for the head part, the root of the north-west extension, and  the east wing were estimated to be  $\sim 120~\mu$G, $\sim 70~\mu$G, and $\sim 32~\mu$G, respectively. 
  The larger field value for  the head part and the lower field value for the wing could be explained by the configuration between the field direction and the UV incident direction (e.g.,  near parallel or perpendicular), considering the results of the numerical modeling of bright-rimmed clouds in quasi-stationary equilibrium after compression by the UV impact \citep{2013ApJ...766...50M}.
  The mass-to-flux ratio normalized by the critical mass-to-flux ratio for the magnetic stability were also calculated to be $\sim 3.0$,  $\sim 1.1$ and $\sim 1.6$ in these three areas, respectively. 
  These values indicate that the head part tends to be most likely supercritical than the other two areas, consistent with the fact that active star formation has already occurred in the head part.
    
  \item We consider the possibility that the two-part-like structure of the cloud have been made from a single elongated cloud by referring the radiation-hydrodynamic simulations of prolate clouds at H\emissiontype{II} boundaries when illuminated by UV radiation from different incident directions \citep{2015MNRAS.450.1017K}. 
We found that the hydrostatic simulations successfully reproduced several structures similar to the IC~1396E/SFO~38 cloud in morphology when the UV incidence direction was oblique to the cloud elongation (i.e.,  head + extension = nose +  filament).
 Although the simulations did not take the magnet field effects into account, but given the weak, initial field strength of the cloud, the early evolution of weakly magnetized prolate clouds could expected to be similar to that of non-magnetized ones, by considering the above 3D MHD simulation results.
\end{enumerate}

\section*{Funding}
 This work was partly supported by Grants-in-Aid for Scientific Research (16H05730) from the Ministry of Education, Culture, Sports, Science and Technology (MEXT) of Japan and the East Asian Observatory.

\begin{ack}
“These observations were obtained by the James Clerk Maxwell Telescope, operated by the East Asian Observatory on behalf of the National Astronomical Observatory of Japan; Academia Sinica Institute of Astronomy and Astrophysics; the Korea Astronomy and Space Science Institute; the National Astronomical Research Institute of Thailand; Center for Astronomical Mega-Science (as well as the National Key R\&D Program of China with No. 2017YFA0402700). Additional funding support is provided by the Science and Technology Facilities Council of the United Kingdom and participating universities and organizations in the United Kingdom and Canada.” 
Additional funds for the construction of SCUBA-2 were provided by the Canada Foundation for Innovation.
The authors wish to recognize and acknowledge the very significant cultural role and reverence that the summit of Maunakea has always had within the indigenous Hawaiian community.  We are most fortunate to have the opportunity to conduct observations from this mountain.
KS and TK would like dedicate this paper to the late Dr. J.~Miao, who was T.~M.~Kinnear's supervisor. 
\end{ack}


\appendix 
\section*{ADF Fitting Results}
\begin{figure*}
\includegraphics[width=12cm]{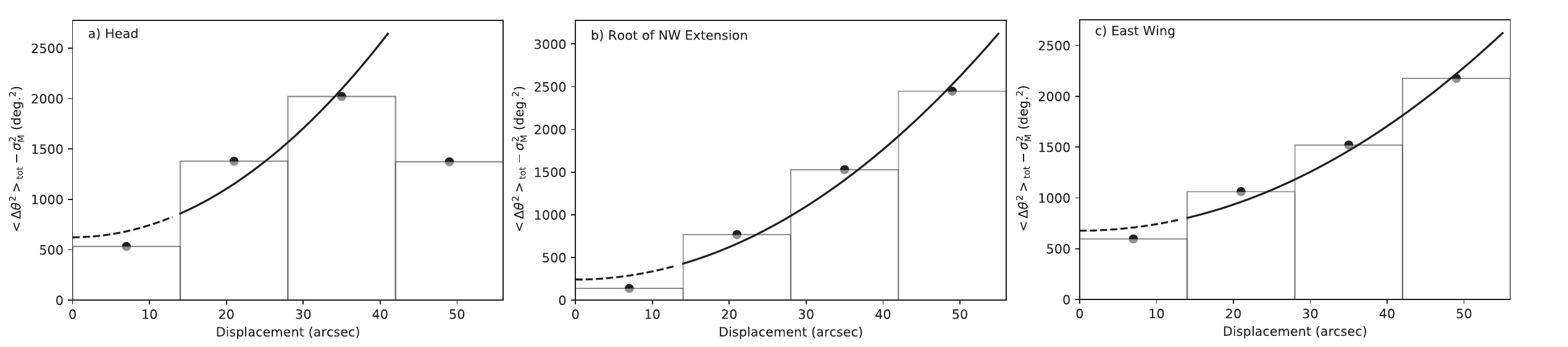}
\caption{Square of the total  measured dispersion function versus displacement for the  three selected areas. 
The square of the measurement  uncertainty $\sigma_\mathrm{M}^2$ is subtracted from the square of the the total measured dispersion function. 
The best-fit power law line is shown by a solid line in each panel and $b^2$ is estimated at the intercept point of the displacement of $l=0$.
{Alt text: Three XY scatter plots with fitting lines.}
}
 \label{fig11}
\end{figure*}

The ADF fitting results in Section 4 are shown in figure \ref{fig11}.


\begin{thebibliography}{}

\bibitem[Andersson, Lazarian, and Vaillancourt(2015)]{2015ARA&A..53..501A}
  Andersson, B.~G., Lazarian, A., \& Vaillancourt, J. E.\ 2015, \araa, 53, 501
\bibitem[Beltr{\'a}n et al.(2004)]{2004A&A...426..941B}
  Beltr{\'a}n, M.~T., Girart, J.~M., Estalella, R., \& Ho, P.~T.~P.\ 2004, \aap, 426, 941
\bibitem[Beltr{\'a}n et al.(2002)]{2002ApJ...573..246B}
  Beltr{\'a}n, M.~T., Girart, J.~M., Estalella, R., Ho, P.~T.~P. \& Palau, A. \ 2002, \apj, 573, 246
\bibitem[Beltr{\'a}n et al.(2012)]{2012A&A...542L..26B}
  Beltr{\'a}n, M.~T., Massi, F., Fontani, F., Codella, C., \& López, R.\ 2012, \aap, 542, L26
\bibitem[Beltr{\'a}n et al.(2009)]{2009A&A...504...97B}
  Beltr{\'a}n, M.~T., Massi, F., L{\'o}pezi, R., Girart, J.~M., \& Estalella, R.\ 2009, \aap, 504, 97
\bibitem[Bertoldi(1989)]{1989ApJ...346..735B}
  Bertoldi, F.\ 1989, \apj, 346, 735
\bibitem[Bertoldi \& McKee(1990)]{1990ApJ...354..529B}
  Bertoldi, F., \& McKee, C.~F.\ 1990, \apj, 354, 529 
\bibitem[Bhatt(1999)]{1999MNRAS.308...40B}
  Bhatt, H.~C.\ 1999, \mnras, 308, 40
\bibitem[Bhatt, Maheswar, and Manoj(2004)]{2004MNRAS.348...83B}
  Bhatt, H.~C., Maheswar, G., \& Manoj, P.\ 2004, \mnras, 348, 83
\bibitem[Bisbas et al.(2009)]{2009A&A...497..649B}
  Bisbas, T. G., W{\"u}nsch, R., Whitworth, A.~P., \& Hubber, D.~A.\ 2009, \aap, 497, 649
\bibitem[Bisbas et al.(2011)]{2011ApJ...736..142B}
  Bisbas, T. G., W{\"u}nsch, R., Whitworth, A.~P., Hubber, D.~A. \& Walch, S.\ 2011, \apj, 736, 142
\bibitem[Chandrasekhar \& Fermi(1953)]{1953ApJ...118..113C}
  Chandrasekhar, S., \& Fermi, E.\ 1953, \apj, 118, 113 
\bibitem[Chapin et al.(2013)]{2013MNRAS.430.2545C}
  Chapin, E.~L., Berry, D.~S., Whitworth, A.~P., Gibb, A.~G. et al.\ 2013, \mnras, 430, 2545
\bibitem[Choudhury, Mookerjea, and Bhatt(2010)]{2010ApJ...717.1067C}
  Choudhury, R., Mookerjea, B., \& Bhatt, H.~C.\ 2010, \apj, 717, 1067
\bibitem[Chuss et al.(2019)]{2019ApJ...872..187C}
 Chuss, D.~T., Andersson, B,~G., Bally, J. et al. \ 2019, \apj, 872, 187
\bibitem[Codella et al.(2001)]{2001A&A...376..271C}
  Codella, C., Bachiller, R., Nisini, B., Saraceno, P., \& Testi, L.\ 2001, \aap, 376, 271
\bibitem[Contreras et al.(2002)]{2002AJ....124.1585C}
  Contreras, M.~E., Sicilia-Aguilar, A., Muzerolle, J., et al.\ 2002, \aj, 124, 1585 
\bibitem[Crampton \& Redman(1975)]{1975AJ.....80..454C}
  Crampton, D., \& Redman, R.~O.\ 1975, \aj, 80, 454
\bibitem[Currie et al.(2014)]{2014ASPC..485..391C}
  Currie, M. J., Berry, D.~S., Jenness, T., et al.\ 2014, in Astronomical Society of the Pacific Conference Series, Vol. 485, Astronomical Data Analysis Software and Systems XXIII, ed. N. Manset \& P. Forshay, 391 
\bibitem[Davis(1951)]{1951PhRv...81..890D}
  Davis, L.\ 1951, Physical Review, 81, 890 
\bibitem[Dempsey et al.(2013)]{2013MNRAS.430.2534D}
  Dempsey, J.~T., Friberg, P., Jenness, T., et al.\ 2013, \mnras, 430, 2534         
\bibitem[Elmegreen(1998)]{1998ASPC..148..150E}
  Elmegreen, B.~G.\ 1998, in Astronomical Society of the Pacific Conference Series, Vol. 148, Origins, ed.\  C.~E. Woodward, J.~M. Shull, \& J. Thronson, Harley A. 150 
\bibitem[Elmegreen(2011)]{2011EAS....51...45E}
  Elmegreen, B.~G.\ 2011, in EAS Publications Series, Vol. 51, EAS Publications Series, ed.\  C. Charbonnel \& T. Montmerle, 45–58 
\bibitem[Felli, Palagi, and Tofani(1992)]{1992A&A...255..293F}
  Felli, M., Palagi, F., \& Tofani, G.\ 1992, \aap, 255, 293
\bibitem[Friberg(2016)]{2016SPIE.9914E..03F}
  Friberg, P., Bastien, P., Berry, D., et al.\ 2016, in Society of 564 Photo-Optical Instrumentation  Engineers (SPIE) Conference Series, Vol. 9914, Millimeter, Submillimeter, and Far-Infrared Detectors and Instrumentation for Astronomy VIII, ed. W. S. Holland \& J. Zmuidzinas, 991403
\bibitem[Fuente et al.(2009)]{2009A&A...507.1475F}
  Fuente, A., Castro-Carrizo, A., Alonso-Albi, T. et al.\ 2009, \aap, 507, 1475
\bibitem[Gaia Collaboration et al.(2018)]{2018A&A...616A...1G}
  Gaia Collaboration, Brown, A.~G.~A., Vallenari, A., et al.\ 2018, \aap, 616, A1
\bibitem[Gaia Collaboration et al.(2021)]{2021A&A...649A...1G}
  Gaia Collaboration, Brown, A.~G.~A., Vallenari, A., et al.\ 2021, \aap, 649, A1 
\bibitem[Getman et al.(2007)]{2007ApJ...654..316G}
  Getman, K.~V., Feigelson, E.~D., Garmire, G., Broos, P., \& Wang, J.\ 2007, \apj, 654, 316 
\bibitem[Grieder(2001)]{2001cre..book.....G}
  Grieder, P.~K.~F.\ 2001, Cosmic Rays at Earth (Elsevier)
\bibitem[Gritschneder et al.(2009)]{2009MNRAS.393...21G}
  Gritschneder, M., Naab, T., Garmire, G., Burkert, A., et al.\ 2009, \mnras, 393, 21 
\bibitem[Henney et al.(2009)]{2009MNRAS.398..157H}
  Henney, W.~J., Arthur, S. J., de Colle, F., \& Mellema, G.\ 2009, \mnras, 398, 157 
\bibitem[Hildebrand et al.(2009)]{2009ApJ...696..567H}
  Hildebrand, R.~H., Kirby, L., Dotson, J. L., Houde, M., \& Vaillancourt, J. E.\ 2009, \apj, 696, 567 
\bibitem[Holland et al.(2013)]{2013MNRAS.430.2513H}
  Holland, W.~S., Bintley, D., Chapin, E. L., et al.\ 2013, \mnras, 430, 2513 
\bibitem[Hollenbach \& Tielens(1999)]{1999RvMP...71..173H}
  Hollenbach, D. J., \& Tielens, A.~G.~G. M.\ 1999, Reviews of Modern Physics, 71, 173
\bibitem[Hora et al.(2008)]{2008PASP..120.1233H}
  Hora, J.~L., Carey, S., Surace, J., et al.\ 2008, \pasp, 120, 1233 
\bibitem[Houde et al.(2009)]{2009ApJ...706.1504H}
  Houde, M., Vaillancourt, J.~E., Hildebrand, R.~H., Chitsazzadeh, S., \& Kirby, L.\ 2009, \apj, 706, 1504
\bibitem[Kessel-Deynet \& Burkert(2003)]{2003MNRAS.338..545K}
  Kessel-Deynet, O., \& Burkert, A.\ 2003, \mnras, 338, 545 
\bibitem[Kinnear et al.(2014)]{2014MNRAS.444.1221K}
  Kinnear, T.~M., Miao, J., White, G.~J., \& Goodwin, S.\ 2014, \mnras, 444, 1221 
\bibitem[Kinnear et al.(2015)]{2015MNRAS.450.1017K}
  Kinnear, T.~M., Miao, J., White, G.~J., Sugitani, K., \& Goodwin, S.\ 2015, \mnras, 450, 1017 
\bibitem[K{\"o}nyves et al.(2010)]{2010A&A...518L.106K}
  K{\"o}nyves, V., Andr{\'e}, P., Men'shchikov, A., et al.\ 2010, \aap, 518, 106 
\bibitem[Kusune et al.(2015)]{2015ApJ...798...60K}
  Kusune, T., Sugitani, K., Miao, J., et al.\ 2015, \apj, 798, 60 
\bibitem[Kwon et al.(2018)]{2018ApJ...859....4K}
  Kwon, J., Doi, Y., Tamura, M., et al.\ 2018, \apj, 859, 4 
\bibitem[Lefloch \& Lazareff(1994)]{1994A&A...289..559L}
  Lefloch, B., \& Lazareff, B. \ 1994, \aap, 289, 559  
 \bibitem[Liu, Zhang, and Qiu(2022)]{2022FrASS...9.3556L}
  Liu, J., Zhang, Q., \& Qiu, K. \ 2022, Frontiers in Astronomy and Space Sciences, vol. 9, id. 943556 
 \bibitem[Liu, Qiu, and Zhang(2022)] {2022ApJ...925...30L}
  Liu, J., Qiu, K., \& Zhang, Q. \ 2022, \apj, 925, 30 
\bibitem[Mackey \& Lim(2010)]{2010MNRAS.403..714M}
  Mackey, J., \& Lim, A.~J. \ 2010, \mnras, 403, 714  
\bibitem[Mackey \& Lim(2011)]{2011MNRAS.412.2079M}
  Mackey, J., \& Lim, A.~J. \ 2011, \mnras, 412, 2079  
\bibitem[Matthews(1979)]{1979A&A....75..345M}
  Matthews, H. I.\ 1979, \aap, 75, 345
\bibitem[Miao et al.(2010)]{2010ApJ...717..658M}
  Miao, J., Sugitani, K., White, G.~J., \& Nelson, R.~P.\ 2010, \apj, 717, 658 
 \bibitem[Miao et al.(2006)]{2006MNRAS.369..143M}
  Miao, J., White, G.~J., Nelson, R.~P., Thompson, M., \& Morgan, L.\ 2006, \mnras, 369, 143 
\bibitem[Miao et al.(2009)]{2009ApJ...692..382M}
  Miao, J., White, G.~J., Thompson, M., \& Nelson, R.~P\ 2009, \apj, 692, 382
\bibitem[Morgan et al.(2004)]{2004A&A...426..535M}
  Morgan, L. K., Thompson, M.~A., Urquhart, J.~S., White, G.~J., \& Miao, J.\ 2004, \aap, 462, 535
\bibitem[Motoyama et al.(2013)]{2013ApJ...766...50M}
  Motoyama, K., Umemoto, T., Shang, H., \& Hasegawa, T.\ 2013, \apj, 766, 50
\bibitem[Nagashima(1999)]{1999sf99.proc..397N}
  Nagashima, C., Nagayama, T., Nakajima, Y., et al.\ 1999, in Star Formation 1999, ed. T. Nakamoto (Nobeyama: Nobeyama Radio Observatory), 397-398  
\bibitem[Nagayama et al.(2003)]{2003SPIE.4841..459N}
  Nagayama, T., Nagashima, C., Nakajima, Y., et al.\ 2003, \procspie, 4841, 459 
 \bibitem[Nakano et al.(2012)]{2012AJ....143...61N}
  Nakano, M., Sugitani, K., Watanabe, M., et al.\ 2012, \aj, 143, 61 
\bibitem[Nakano \& Nakamura(1978)]{1978PASJ...30..671N}
  Nakano, T., \& Nakamura, T. \ 1978, \pasj, 30, 67  
\bibitem[Neri et al.(2007)]{2007A&A...468L..33N}
  Neri, R., Fuente, A., Ceccarelli, C., et al.\ 2007, \aap, 468, L33 
\bibitem[Nisini et al.(2001)]{2001A&A...376..553N}
  Nisini, B., Massi, F., Vitali, F., et al.\ 2001, \aap, 376, 553
\bibitem[Ogura,  Sugitani, and Pickles(2002)]{2002AJ....123.2597O}
  Ogura, K., Sugitani, K., \& Pickles, A.\ 2002, \aj, 123, 2597 
 \bibitem[Ostriker,  Stone, and Gammie(2001)]{2001ApJ...546..980O}
  Ostriker, E. C., Stone, J. M., \& Gammie, C. F.\ 2001, \apj, 546, 980 
\bibitem[Patel et al.(1995)]{1995ApJ...447..721P}
  Patel, N.~A., Goldsmith, P.~F., Snell, R.~L., Hezel, T., \& Xie, T.\ 1995, \apj, 447, 721
\bibitem[Patel et al.(2000)]{2000ApJ...538..268P}
  Patel, N.~A., Greenhill, L.~J., Herrnstein, J., et al.\ 2000, \apj, 538, 268
\bibitem[Pattle et al.(2018)]{2018ApJ...860L...6P}
  Pattle, K., Ward-Thompson, D., Hasegawa, T., et al.\ 2018, \apjl, 860, L6 
\bibitem[Pedlar(1980)]{1980MNRAS.192..179P}
  Pedlar, A.\ 1980, \mnras, 192, 179 
\bibitem[Pottasch(1956)]{1956BAN....13...77P}
  Pottasch, S.~R.\ 1956, \bain, 13, 77 
\bibitem[Planck Collaboration et al.(2015)]{2015A&A...576A.104P}
  Planck Collaboration et al.\ 2015, \aap, 576, A104
\bibitem[Rathborne et al.(2002)]{2002MNRAS.331...85R}
  Rathborne, J. M., Burton, M. G., Brooks, K. J., et al.\ 2002, \mnras, 331, 85 
\bibitem[Reipurth et al.(2003)]{2003ApJ...593L..47R}
  Reipurth, B., Armond, T., Raga, A., \& Bally, J.\ 2003 , \apjl, 593, L47 
\bibitem[Sandford, Whitaker, and Klein(1982)]{1982ApJ...260..183S}
  Sandford, M.~T., Whitaker, R.~W., \& Klein, R.~I.\ 1982, \apj, 260, 185   
\bibitem[Santos et al.(2014)]{2014ApJ...783....1S}
  Santos, F.~P., Franco, G.~A.~P., Roman-Lopes, A., Reis, W., \& Rom{\'a}n-Z{\'u}{\~n}iga, C.~G.\ 2014, \apj, 783, 1 
\bibitem[Saraceno et al.(1996)]{1996A&A...315L.293S}
  Saraceno, P., Ceccarelli, C., Clegg, P., et al.\ 1996, \aap, 315, L293 
\bibitem[Serabyn, Guesten, and Mundy(1983)]{1993ApJ...404..247S}
  Serabyn, E., Guesten, R., \& Mundy, L.\ 1993, \apj, 404, 247
\bibitem[Sicilia-Aguilar et al.(2019)]{2019A&A...622A.118S}
  Sicilia-Aguilar, A., Patel, N., Fang, M., et al.\ 2019, \aap, 622, 118 
\bibitem[Silverberg et al.(2021)]{2021AJ....162..279S}
  Silverberg, S.~M., Günther, H.~M., Kim, J.~S., Principe, D.~A., \& Wolk, S.~J.\ 2021, \aj, 162, 279
\bibitem[Slysh et al.(1999)]{1999ApJ...526..236S}
  Slysh, V. I., Val'tts, I. E., Migenes, V., et al.\ 1999, \apj, 526, 236 
\bibitem[Soam et al.(2013)]{2013MNRAS.432.1502S}
  Soam, A., Maheswar, G., Bhatt, H.~C., Lee, C.~W., \& Ramaprakash, A.~N.\ 2013, \mnras, 432, 150
\bibitem[Soam et al.(2017)]{2017MNRAS.465..559S}
  Soam, A., Maheswar, G., Lee, C.~W., Neha, S., \& Andersson, B.~G.\ 2017, \mnras, 465, 559
 \bibitem[Soam et al.(2017)]{2018MNRAS.476.4782S}
  Soam, A., Maheswar, G., Lee, C.~W., Neha, S., \& Kim, K.-T.\ 2018, \mnras, 476, 4782
\bibitem[Sridharan,  Bhatt, and Rajagopal(1996)]{1996MNRAS.279.1191S}
  Sridharan, T. K., Bhatt, H. C., \& Rajagopal, J.\ 1996, \mnras, 279, 1191 
\bibitem[Sugitani et al.(1989)]{1989ApJ...342L..87S}
  Sugitani, K., Fukui, Y., Mizuni, A., \& Ohashi, N.\ 1989, \apjl, 342, L87 
\bibitem[Sugitani,  Fukui, and Ogura(1991)]{1991ApJS...77...59S}
  Sugitani, K., Fukui, Y., \& Ogura, K.\ 1991, \apjs, 77, 59 
 \bibitem[Sugitani et al.(2000)]{2000AJ....119..323S}
  Sugitani, K., Matsuo, H., Nakano, M., Tamura, M., \& Ogura, K.\ 2000, \aj, 119, 323
\bibitem[Sugitani et al.(2002)]{2002aprm.conf..213S}
  Sugitani, K., Fukui, Y., \& Ogura, K.\ 2002, in 8th Asian-Pacific Regional Meeting, Volume II, ed. S.~Ikeuchi, J.~Hearnshaw, \& T.~Hanawa, 213-214
\bibitem[Sugitani et al.(2007)]{2007PASJ...59..507S}
  Sugitani, K., Watanabe, M., Tamura, M., et al.\ 2007, \pasj, 59, 507 
\bibitem[Targon et al.(2011)]{2011ApJ...743...54T}
  Targon, C.~G., Rodrigues, C.~V., Cerqueira, A.~H., \& Hickel, G.~R.\ 2011, \apj, 743, 54 
\bibitem[Tofani et al.(1955)]{1995A&AS..112..299T}
  Tofani, G., Felli, M., Taylor, G.~B., \& Hunter, T.~R.\ 1995, \aaps, 112, 299 
\bibitem[Urquhart et al.(2003)]{2003A&A...409..193U}
  Urquhart, J.~S., White, G.~J., Pilbratt, G.~L., \& Fridlund, C.~V.~M.\ 2003, \aap, 409, 193 
\bibitem[Walborn \& Panek(1984)]{1984ApJ...286..718W}
  Walborn, N.~R., \& Panek, R.~J.\ 1984, \apj, 286, 718  
\bibitem[Weikard et al.(1996)]{1996A&A...309..581W}
  Weikard, H., Wouterloot, J.~G.~A., Castets, A., Winnewisser, G., \& Sugitani, K.\ 1996, \aap, 309, 581 
\bibitem[Werner et al.(2004)]{2004ApJS..154....1W}
  Werner, M. W., Roellig, T. L., Low, F. J., et al.\ 2004, \apjs, 154, 1 
\bibitem[Wilking et al.(1993)]{1993AJ....106..250W}
  Wilking, B., Mundy, L., McMullin, J., Hezel, T., \& Keene, J.\ 1993, \aj, 106, 250 
\bibitem[Wilking, Blackwell, and Mundy(1990)]{1990AJ....100..758W}
  Wilking, B.~A., Blackwell, J.~H., \& Mundy, L.~G.\ 1990, \aj, 100, 758         
\bibitem[Williams, Ward-Thompson, and Whitworth(2001)]{2001MNRAS.327..788W}
  Williams, R.~J.~R., Ward-Thompson, D., \& Whitworth, A.~P.\ 2001, \mnras , 327, 788   
\bibitem[Wright et al.(2010)]{2010AJ....140.1868W}
  Wright, E. L., Eisenhardt, P. R. M., Mainzer, A. K., et al.\ 2010, \aj, 140, 1868

\end{thebibliography}

\end{document}